\documentclass[twocolumn,english,showpacs,aps,prb]{revtex4}

\usepackage[normalem]{ulem}

\usepackage[latin9]{inputenc}
\usepackage{graphicx}
\usepackage{amsmath}
\usepackage{amssymb}
\usepackage{bbm}
\usepackage{xspace}
\usepackage{color}
\usepackage{footnote}
\usepackage{comment}
\usepackage{hyperref}
\usepackage{subfigure}

\definecolor{dgreen}{rgb}{0.0,0.5,0.0}

\definecolor{orange}{RGB}{252,77,6}
\definecolor{gray}{RGB}{50,50,50}
\definecolor{brown}{RGB}{200,127,50}
\definecolor{blue}{RGB}{00,000,100}
\definecolor{blue2}{RGB}{00,000,250}
\definecolor{green1}{RGB}{00,100,00}
\definecolor{green2}{RGB}{00,150,00}
\definecolor{green3}{RGB}{00,200,00}
\definecolor{green4}{RGB}{00,250,00}

\newcommand{\fig}[1]{Fig.~\thinspace{}\ref{#1}}

\newcommand{\eq}[1]{Eq.\thinspace{}(\ref{#1})}

\newcommand{\beq}{\begin{equation}}
\newcommand{\eeq}{\end{equation}}

\newcommand{\vv}[1]{{\boldsymbol{#1}}_\perp}

\newcommand{\Imm}{{\rm Im\:}}

\newcommand{\avg}[1]{\langle #1 \rangle}

\begin{document}

\title{Thermoelectric properties of a strongly correlated layer}

\author{Irakli Titvinidze}

\email{irakli.titvinidze@tugraz.at}


\affiliation{Institute of Theoretical and Computational Physics, Graz University
of Technology, 8010 Graz, Austria}

\author{Antonius Dorda}

\affiliation{Institute of Theoretical and Computational Physics, Graz University
of Technology, 8010 Graz, Austria}

\author{Wolfgang von der Linden}

\affiliation{Institute of Theoretical and Computational Physics, Graz University
of Technology, 8010 Graz, Austria}

\author{Enrico Arrigoni}

\affiliation{Institute of Theoretical and Computational Physics, Graz University
of Technology, 8010 Graz, Austria}


\begin{abstract} 
In this paper we investigate  the effect of strong electronic interactions  on the thermoelectric properties of a simple generic system, consisting of a single correlated layer sandwiched between two metallic leads. Results will be given for the linear response regime as well as beyond, for which a full nonequilibrium many-body calculation is performed, based on nonequilibrium dynamical mean-field theory (DMFT). 
As impurity solver we use the auxiliary master equation approach, which addresses the impurity problem within a finite auxiliary system consisting of a correlated impurity, a small number of uncorrelated bath sites and  Markovian environments.
For the linear response regime, results will be presented for the Seebeck coefficient, the electrical conductance and the electronic  contribution to the thermal conductance. 
Beyond linear response, i.e. for finite  differences  in the temperatures and/or the bias voltages in the leads, we study the dependence of the current on various model parameters, such as gate voltage  and Hubbard interaction of the central layer. We will give a detailed parameter study as far as the thermoelectric efficiency is concerned. We find that strong correlations can indeed increase the thermopower of the device. In addition, some general theoretical requirements for an efficient thermoelectric device will be given.
\end{abstract}

\ifx\clength\undefined
\maketitle
\else
\nocomm
\fi

\begin{figure}[t]
 \includegraphics[width=\columnwidth]{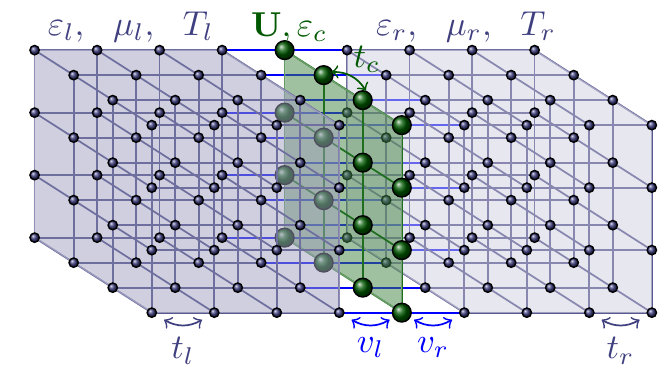}
\caption{(Color online) Schematic representation of the investigated heterostructure, consisting of a single correlated layer (dark green) sandwiched between two semi-infinite metallic leads (grey).  
$U$, $\varepsilon_c$ and $t_c$ are local Hubbard interaction, onsite energy and the hopping amplitudes between neighboring sites of the correlated layer, correspondingly. Hybridization between the correlated 
layer and the left (right) lead is $v_l$ ($v_r$). The hopping between neighboring sites for the left (right) lead is $t_l$ ($t_r$). $\varepsilon_l$ ($\varepsilon_r$) and $\mu_l$ ($\mu_r$) are onsite energy and 
chemical potential for left (right) lead. The temperature of the left and right leads are correspondingly $T_l$ and $T_r$.}
\label{schematicp}
\end{figure}

\section{Introduction}

Thermoelectric devices on the nano scale are an interesting topic and  research object, not only from the pure scientific point of view, but much more so  for   technological reasons. All aspects of thermoelectric devices are nicely reviewed by Benenti et al.\cite{be.ca.13u, be.ca.17}. In particular, the efficient harvesting of the thermal energy is of great socio-economical importance. It depends on the physical properties of the given material and in the linear response regime it can be characterized by the dimensionless quantity, which is called thermoelectric figure of merit 
\begin{align}
\label{Figure_of_merit}
ZT&=\frac{S^{2}}{L}\\
L &=\frac{\kappa}{\sigma T} \;,
\label{Lorenz_ratio}
\end{align}
which is expressed by the Seebeck coefficient $S$ and the Lorenz number $L$. Here $T$ is the temperature, $\sigma$ is the electrical conductance and $\kappa$ is the thermal conductance. Any inelastic scattering  will contribute to the thermal conductivity. However, for simplicity in the present work we consider only the electrical contribution $\kappa=\kappa_e$. A high value of the figure of merit is required to obtain high efficiency.\cite{be.ca.13u, ki.za.13} In spite of the many theoretical and experimental efforts\cite{ha.ta.02, ch.pr.09, zh.dr.11, maju.04, po.ha.08, ve.si.01, ar.ku.08, hs.lo.04, be.nu.02, sa.ma.96, pa.bo.13, pa.bo.13a, sn.to.08, dr.ch.07, du.ve.11, vi.sh.10, so.ch.09,shak.11, do.ga.16, fr.zl.07} most efficient actual devices still operate at $ZT \lesssim 1$.\cite{be.ca.13u} In this paper, we will study the impact of strong correlations on the thermoelectric properties of layered systems. It will turn out that the  Lorenz ratio in these systems is close to the universal Wiedemann-Franz value $L_{0}$. Therefore, the figure of merit is solely governed by the Seebeck coefficient. The system we studied numerically,  is composed of a single correlated layer sandwiched between two metallic leads. We will present results  in and beyond the linear response regime. In the linear response regime we study the behavior of the Seebeck coefficient, electrical conductance and the electronic contribution to the thermal conductance, while in the non-linear regime we will address the electrical current $J$ and the corresponding power $P=J\cdot \Phi$, with $\Phi$ being the applied voltage.

To investigate the behavior of a many-body  system out of equilibrium we use dynamical mean-field theory (DMFT)\cite{ge.ko.96, Voll.10, me.vo.89} combined  with the nonequilibrium Green's function approach originally proposed by Kubo\cite{kubo.57}, Schwinger\cite{schw.61}, Kadanoff, Baym\cite{ba.ka.61, kad.baym} and Keldysh\cite{keld.65}. DMFT is  a comprehensive, thermodynamically consistent and non-perturbative scheme. It is one of the powerful methods to investigate high-dimensional correlated systems and becomes exact in infinite dimensions. The sole approximation is the assumption of a local self-energy. In order to determine it,  the original lattice  problem is mapped onto a single impurity Anderson model (SIAM)\cite{ande.61}, which implies a self-consistency cycle. To solve the impurity problem, particularly in 
the nonequilibrium case, we used a recently developed auxiliary master equation approach (AMEA)\cite{ar.kn.13,do.nu.14,ti.do.15, do.so.17, do.ga.16}. The latter directly treats the nonequilibrium steady-state situation. 
However, AMEA is not restricted to the nonequilibrium situation, it can very efficiently be used  in the equilibrium case as well. AMEA treats the impurity problem within an auxiliary system consisting of the correlated impurity, a small number of uncorrelated bath sites and two Markovian environments, which are encoded in a Lindblad equation.

The paper is organized as follows: In Sec.~\ref{Model} we introduce the Hamiltonian of the system. In Sec.~\ref{Method} we briefly discuss the nonequilibrium DMFT approach together with AMEA, while in 
Sec.~\ref{Thermoelectric_properties} we define important thermoelectric properties which we use in our calculations. Afterwards in Sec.~\ref{Results} we present our results. Sec. \ref{Results_linear_response} is dedicated  to the linear response results, while in Sec.~\ref{Results_temperature_gradient} we present our results in presence of the large temperature difference between the leads. 
Finally, in Sec.~\ref{Conclusions} the results are summarized.

\section{Model and Method}\label{Model_Method}

\subsection{Model}\label{Model}

We consider a model system consisting of a single correlated infinite mono-layer ($z=0$) sandwiched between two metallic leads ($z<0$ and $z>0$), which are semi-infinite (see Fig.~\ref{schematicp}). The system is translationally invariant in the $xy$ plane (parallel to the correlated layer). The Hamiltonian reads
\begin{eqnarray}
\label{Hamiltonian}
&&\hspace{-0.75cm}{\cal H}=
-\hspace{-0.5cm}\sum_{z, \langle {\bf r}^{\phantom\dagger}_{\perp},{\bf r}'_{\perp}\rangle, \sigma}\hspace{-0.5cm}t_{z} c_{z,{\bf r}_{\perp},\sigma}^\dagger c_{z,{\bf r}'_{\perp},\sigma}^{\phantom\dagger} 
-\hspace{-0.45cm}\sum_{\langle z, z'\rangle, {\bf r}_{\perp}, \sigma}\hspace{-0.45cm} t_{zz'} c_{z,{\bf r}_{\perp},\sigma}^\dagger c_{z',{\bf r}_{\perp},\sigma}^{\phantom\dagger} \\
&&\hspace{-0.3cm}+U\hspace{-0.05cm}\sum_{{\bf r}_{\perp}}\hspace{-0.05cm}\big(n_{0,{\bf r}_{\perp},\uparrow}-\frac{1}{2}\big)\big(n_{0,{\bf r}_{\perp},\downarrow}-\frac{1}{2}\big) \nonumber 
+\hspace{-0.25cm}\sum_{z,{\bf r}_{\perp},\sigma}\hspace{-0.2cm}\varepsilon_z n_{z,{\bf r}_{\perp},\sigma} \, .
\end{eqnarray}
Here $c_{z,{\bf r}_{\perp},\sigma}^\dagger$ creates an electron at site ${\bf r}_{\perp}$  of  layer $z$ with spin ${\sigma}$ and $n_{z, {\bf r}_{\perp}, \sigma}=c_{z, {\bf r}_{\perp}, \sigma}^\dagger c_{z, {\bf r}_{\perp}, \sigma}^{\phantom\dagger}$ denotes the corresponding occupation-number operator. $\langle z, z'\rangle$ stands for neighboring layers and $\langle {\bf r}^{\phantom\dagger}_{\perp},{\bf r}'_{\perp}\rangle$ stands for neighboring sites within a layer. 

The first two terms of the Hamiltonian \eqref{Hamiltonian} describe nearest-neighbor intra-layer and inter-layer hoppings, with hopping amplitudes $t_{z}$ and $t_{zz'}$, respectively. We consider identical nearest-neighbor hopping parameters within the leads. $t_{z}=t_{zz'}=t_l$ for the left lead ($z,z' <0 $) and $t_{z}=t_{zz'}=t_r$ for the right lead ($z,z' > 0$). The hopping amplitude inside the correlated layer is $t_0=t_c$, while the hybridization between left (right) lead and correlated layer $t_{-1,0}=t_{0,-1}=v_{l}$ ($t_{0,1}=t_{1,0}=v_{r}$). 
The third term introduces the local Hubbard interactions $U$, which is nonzero only for the correlated layer ($z=0$). The last term describes the onsite energies. For the correlated layer $\varepsilon_{z=0} \equiv \varepsilon_c$, and for the left and right lead $\varepsilon_{z<0} \equiv \varepsilon_l$ and $\varepsilon_{z>0}\equiv \varepsilon_r$, respectively.

The nonequilibrium situation is obtained by applying a bias voltage $\Phi$ and/or temperature difference $\Delta T=T_l-T_r$, between the leads. Here $T_l$ and $T_r$ is the temperature of the left and right lead, respectively.
The usual way to treat such steady state situation (see, e.g. ~\cite{okam.07,okam.08}) is to start  at $t=-\infty$  with three decoupled ($v_l=v_r=0$)  systems consisting of the two leads plus the central region
which are separately in equilibrium at different chemical potentials and temperatures.
Then one switches on the hybridisation and waits until a steady state is reached. 
A bias voltage $\Phi$ is obtained by shifting both the onsite energies $\varepsilon_{l/r}$ of the left and right leads as well as their initial chemical potentials $\mu_{l/r}$ 
by  $\pm |e|\Phi/2$, respectively, i.e. 
$\varepsilon_{l}=\mu_l= |e| \Phi/2$, $\varepsilon_{r}=\mu_r= -|e| \Phi/2$.
Of course, this neglects the long-range Coulomb interaction, which could be added explicitly, for example, by solving self-consistently   the Poisson equation, 
(see, e.g. \cite{fr.zl.07, ok.mi.04, ch.fr.07, ha.fr.12, ba.es.16, ch.ha.13, le.do.06, le.do.07}). 
One major effect of the long-range Coulomb interaction is a voltage drop across the central region. The approximation made here is to assume that the voltage drop takes place only in the central region, which corresponds to the limit  of infinite Coulomb screening.

In the following,  we use units in which $t_c =1, \hbar =1, k_{B} = 1, |e| = 1$, and $a= 1$. The latter is the distance between neighboring sites of the simple cubic lattice. Thus, the current (density) is expressed in units of
\begin{align}\label{eq:j0}
j_{0}&=\frac{|e|}{\hbar a^{2}} \, .
\end{align}

\subsection{Method}\label{Method}

Here we give a brief overview of the nonequilibrium DMFT approach\cite{sc.mo.02u, fr.tu.06, free.08, jo.fr.08, ec.ko.09, okam.07, okam.08} and the auxiliary-master-equation approach 
(AMEA)\cite{ar.kn.13,do.nu.14,ti.do.15, do.so.17}, employed to solve the intrinsic impurity problem. For more details we refer to Refs.~\onlinecite{ar.kn.13, do.nu.14, ti.do.15, ti.do.16, do.ti.16}\, . To describe the behavior of the steady state it is  convenient to work in the Keldysh Green's function formalism~\cite{kad.baym, schw.61, keld.65, ha.ja, ra.sm.86} and introduce $2\times2$ block Green's functions
\begin{equation}
\underline G =\left(
\begin{array}{cc}
G^R & G^K \\
0   & G^A 
\end{array}
\right) \,,
\label{eq:KeldyshGF}
\end{equation}
which we denote by an underscore. $G^R$, $G^K$ and $G^A$ are the retarded, Keldysh, and advanced Green's functions. $G^A$ and $G^R$ are related via $G^A=(G^R)^\dagger$. While $G^K$ can only be obtained from  $G^{R}$ in the equilibrium case via the fluctuation dissipation theorem\cite{ha.ja}, out of equilibrium $G^{K}$ is independent  of $G^{R}$ and it needs to be determined separately, which requires the Keldysh Green's function formalism.

In the steady state,  the two-time Green's functions only depend on the time-difference
and it is convenient to switch to the frequency domain $\omega$. Furthermore, due to  translational invariance along the $xy$-plane we introduce the momentum variable ${\vv  k}=(k_x,k_y)$. The Green's function of the correlated layer is given in terms of Dyson's equation by
\begin{equation}
{\underline G}^{-1}_{c}\hspace{-0.05cm}(\omega, {\vv  k}) ={\underline g}^{-1}_c(\omega, {\vv  k}) - 
\hspace{-0.15cm}\sum_{\alpha=l,r}\hspace{-0.15cm}v_\alpha^2~{\underline g}_\alpha(\omega, {\vv  k}) - {\underline{\Sigma}}(\omega) \,,
\label{eq:Dyson}
\end{equation}
where ${\underline g}_{\alpha}(\omega, {\vv  k})$ denotes the non-interacting, decoupled ($v_l=v_r=0$) equilibrium Green's functions for the correlated layer ($\alpha=c$) and for the edge layer of the left ($l$) and the right ($r$) lead.\cite{Green_g}

The retarded part of the equilibrium Green's functions can be expressed by
\begin{eqnarray}
\label{gcR}
g_c^R(\omega, {\vv  k})&=&\frac{1}{\omega^+ - \varepsilon_c - t_{c} \varepsilon({\vv  k})} \,\\
\label{galphaR}
g_{\alpha=l,r}^R(\omega, {\vv  k})&=&\frac{\omega - \varepsilon_\alpha - t_{\alpha}\varepsilon({\vv  k})}{2t_\alpha^2} \nonumber \\
              &-& i\frac{\sqrt{4t_\alpha^2-(\omega - \varepsilon_\alpha - t_\alpha \varepsilon({\vv  k}))^2}}{2t_\alpha^2} \, 
\end{eqnarray}
with $\varepsilon({\vv  k})=-2(\cos k_x + \cos k_y)$. The sign of the square-root for negative argument in Eq.~\eqref{galphaR} must be chosen such that the Green's function has the correct 
$1/\omega$ behavior for $|\omega|\to \infty$. Since the disconnected leads are separately in equilibrium, we can obtain their Keldysh components from the retarded ones via the fluctuation dissipation theorem\cite{Green_g, ha.ja}
\begin{equation}
g_{\alpha=l,r}^K(\omega, {\vv  k})= 2i(1-2f_\alpha(\omega))\;\Imm g_{\alpha}^R(\omega,{\vv  k}) \, ,
\label{galphaK}
\end{equation}
where $f_\alpha(\omega)$ is the Fermi distribution for chemical potential $\mu_\alpha$ and temperature $T_\alpha$. For the non-interacting isolated central layer the inverse Keldysh Green's function 
$[{\underline g}_{c}(\omega, {\vv  k})]^{-1}$ is proportional to $0^{+}$, an infinitesimal imaginary part, and can be neglected in the final steady state in which the leads and the layers are connected to each other.

${\underline{\Sigma}}(\omega)$ in \eq{eq:Dyson} stands for  the self-energy, which in DMFT is  approximated by a local and therefore momentum ($\bf k$) independent quantity. It will be determined self-consistently by solving the SIAM problem with the same local parameters $U$ and $\varepsilon_c$ as the original model. More specifically we employ the auxiliary master equation approach
(AMEA)\cite{ar.kn.13, do.nu.14,ti.do.15, do.so.17} to determine the self-energy $\Sigma(\omega)$ in the equilibrium and nonequilibrium case.
The key point of AMEA is to map the infinite SIAM problem to an auxiliary one with a finite number of bath sites $N_b$ and two Markovian environments (sink and reservoir), which are crucial to achieve a steady state in a finite system. The resulting open quantum system is described by a Lindblad equation. Its parameters are, however, not taken from a Born-Markov approximation~\cite{br.pe, scha, nu.do.15} but are used as fit parameters to optimally reproduce the physical hybridization function
\begin{equation}
\underline{\Delta}_\mathrm{ph}(\omega)=\underline{g}_0^{-1}(\omega) - \underline{G}^{-1}_{\rm loc}(\omega)-\underline{\Sigma}(\omega)\,,  
\label{eq:D_ph}
\end{equation}
by the hybridization function $\underline{\Delta}_\mathrm{aux}(\omega)$,
obtained in the auxiliary system. Here $\underline{g}_0^{-1}(\omega)$ is the non-interacting Green's function of the disconnected impurity and the local Green's 
function is obtained by
\begin{equation}
\underline{G}_{\rm loc}(\omega)=\int\limits_{\rm BZ} \frac{d{\vv  k}}{(2\pi)^2}\underline{G}_{c}(\omega,{\vv  k}) \,.
\label{eq:G_loc}
\end{equation}
Clearly, the  accuracy of our impurity solver increases exponentially with the number of bath sites $N_b$ and becomes exponentially exact in the limit $N_b \rightarrow \infty$.\cite{do.so.17, sc.go.16} In practice $N_b=4,6$ is sufficient to obtain reliable results for the current. 
The self consistence cycle proceeds as follows: Starting out with an initial guess for the self energy, we determine the physical hybridization function
in \eq{eq:D_ph}. Next the Lindblad parameters of the auxiliary system without interaction are adjusted by minimizing the misfit between the hybridization function of the auxiliary system and that of the physical system. 
Once the Lindblad  parameters are determined, the 
auxiliary nonequilibrium many-body system is solved 
by standard numerical techniques, which defines the self energy for the next iteration. This procedure is repeated until convergence is reached. For more details see Refs.~\onlinecite{ar.kn.13, do.nu.14, ti.do.15, ti.do.16}.

\subsection{Thermoelectric properties}\label{Thermoelectric_properties}

Once the Green's function for the central layer is determined, one can calculate the desired thermoelectric properties. Here we shortly overview the corresponding expressions, which we need in the present work and discuss some of their generic features.

The electrical current density ($J_\text{c}$) and the energy-current density ($J_\text{en}$) through the central layer in case of the linear response regime or zero bias voltage $\Phi=0$ but with finite temperature difference between leads are given by\cite{ha.ja, jauh} (see Appendix \ref{Appendix:Meir-Wingreen})
\begin{eqnarray}\label{Current}
&& \hspace{-0.75cm}J_\text{c/en}= -2\int\limits_{-\infty}^\infty \hspace{-0.1cm}\frac{d\omega}{2\pi}
{\cal T}(\omega)
\left(f_l(\omega)-f_r(\omega)\right)\;\zeta_\text{c/en}(\omega)\;.
\end{eqnarray}
This expression is valid if $\Imm g_l^R(\omega,{\vv  k}) =\lambda \Imm g_r^R(\omega,{\vv  k})$, where $\lambda$ does not depend on $\omega$ and $\bf k_{\perp}$. Otherwise one has to use more general expression given by the Meir-Wingreen formula (see e.g. Ref.\cite{ha.ja, me.wi.92, jauh, re.zh.12, ti.do.15}).

For the electric current density (energy current density) we have $\zeta_\text{c}=1$ ( $\zeta_\text{en}=\omega$).
The transmission function is given by
\begin{align}
{\cal T}(\omega) &= \frac{1}{N_{\perp}}\sum_{k}^{BZ} 
 {\cal T}(\omega,{\vv  k})\\
{\cal T}(\omega,{\vv  k})&=\frac{2\pi\gamma_l(\omega,{\vv  k})\gamma_r(\omega,{\vv  k})}{\gamma_l(\omega,{\vv  k})+\gamma_r(\omega,{\vv  k})}A(\omega,{\vv  k}) \, .
\label{eq:Twk}
\end{align}
The sum runs over the ${\vv k}$-vectors of  the first Brillouin zone of the 
2D  central layer and $N_{\perp}$ is the number of the corresponding vectors.
The transmission function depends on 
the spectral density of the central layer
\begin{eqnarray}
\label{Spectral_function}
A(\omega,{\vv  k})=-\frac{1}{\pi}\Imm G^R_{c}(\omega,{\vv  k})\;,
\end{eqnarray}
and the imaginary part of the surface Green's functions of the leads via
\begin{align}\label{eq:gamma}
\gamma_\alpha(\omega,{\vv  k})=-2v_\alpha^2 \Imm g_\alpha^R(\omega,{\vv  k}).
\end{align}
Obviously, the ${\vv k}$-dependence of  both quantities 
enters through
\begin{align}\label{eq:varepsilon}
\varepsilon ({\vv k}) &= -2 \big(
\cos k_{x} + \cos k_{y}
\big)\;,
\end{align}
with $\gamma_{\alpha}(\omega,{\vv k}) = \tilde\gamma_{\alpha}(\omega,t_{\alpha} \varepsilon({\vv k}))$ and $A(\omega,{\vv k}) =\tilde A(\omega,t_{c} \varepsilon({\vv k}))$. 
Therefore, by replacing $\gamma_{\alpha}(\omega,{\vv k})$ and $A(\omega,{\vv k})$ in \eq{eq:Twk} by $\tilde\gamma_{\alpha}(\omega,t_{\alpha}\varepsilon)$ and $\tilde t A(\omega,t_{c} \varepsilon)$, respectively, we obtain
\begin{align*}
{\cal T}(\omega) &= \int\limits_{-\infty}^{\infty} 
\rho_{2D}(\varepsilon)
\tilde{\cal T}(\omega,\varepsilon)  \;d\varepsilon\;,
\end{align*}
where $\rho_{2D}$ is the 2D density of states corresponding to the dispersion relation $\varepsilon({\vv k})$, which can be expressed in terms of the complete elliptic integral of the first kind ${\cal K}(x)$\cite{Elliptic_K}
as 
\begin{align}\label{eq:2Ddos}
\rho_{2D}(\varepsilon) &= \frac{\theta(4-|\varepsilon|)}{2\pi^{2}}\;{\cal K}\big(1-\big(\frac{\varepsilon}{4}\big)^{2}\big)\;.
\end{align}

As outlined before,  the bias voltage enters the chemical  potential of the leads and the onsite energies, both are given by $\Phi/2$ ($-\Phi/2$ ) for the left (right) lead. However, the change of the onsite energies enters only in second order and does not contribute to the linear response expressions. Straightforward Taylor expansion then yields\cite{re.zh.12, ki.za.13, be.ca.13u}
\begin{align}
\label{J_Phi_Delta_T}
J_\text{el} &= \sigma  \Phi + \sigma  S \Delta T  \;.\\
\label{JQ_Phi_Dela_T}
J_\text{en} &= S T J_\text{el} + \kappa_{e} \Delta T  \;.
\end{align}
The linear response coefficients are the electric conductivity
\begin{equation}
\label{Conductance}
\sigma =\left.\frac{\partial J}{\partial \Phi}\right|_{\Delta T=0} = I_0\;,
\end{equation}
 the Seebeck coefficient
\begin{equation}
\label{Seebeck}
S=-\left.\frac{\partial \Phi}{\partial \Delta T}\right|_{J_\text{el}=0} =-\frac{I_1}{I_0}\beta\;,
\end{equation}
and the electronic contribution to the 
 thermal conductance 
\begin{equation}
\label{ThermalConductance}
\kappa_e=\left.\frac{\partial J_\text{en}}{\partial \Delta T}\right|_{J_\text{el}=0} 
= \sigma\;\left[\frac{I_2}{I_{0}}-\frac{I_1^2}{I_0^{2}}\right]\;\beta\;.
\end{equation}

These expressions are based on the Lorenz integral
\begin{eqnarray}
\label{I_m}
\hspace{-0.25cm}
I_m=  2\int\limits_{-\infty}^\infty \hspace{-0.225cm}\frac{d\omega}{2\pi}
{\cal T}(\omega) \left|f'(\omega)\right| \;\omega^m \, ,
\end{eqnarray}
with $f(\omega) =1/(\exp(\beta\omega)+1)$.

In addition we also calculated the Lorenz ratio (see Eq.~\eqref{Lorenz_ratio}), 
which in the linear response regime can be expressed as
\begin{align*}
L &= \frac{\kappa_{e}}{\sigma T}
=\left[\frac{I_2}{I_{0}}-\frac{I_1^2}{I_0^{2}}\right]\;\beta^{2}\;.
\end{align*}
We will check the validity of the Wiedemann-Franz law
\begin{eqnarray}
\label{WFL}
L=L_0=\frac{\pi^2}{3} \, 
\end{eqnarray}
in the presence of van Hove singularities and interacting electrons. The Wiedemann-Franz law is assumed to be valid for a free electron gas with elastic scattering. 
By definition, the transmission function ${\cal T}(\omega)$ is non-negative and so is 
\begin{align}
|f'(\omega)| = \frac{\beta}{4  \cosh^{2}(\frac{\beta\omega}{2})}\;.
\end{align}
The quantities $I_{m}/I_{0}$ can therefore be considered as the $m$-th moment 
\begin{equation}
\label{omegam}
\avg{ \omega^{m}} =\int d\omega \rho(\omega) \;\omega^{m}\;. 
\end{equation}
of the probability distribution function
\begin{align*}
\rho(\omega) &= \frac{1}{Z}\frac{{\cal T}(\omega)}{\cosh^{2}(\frac{\beta\omega}{2})}\;,
\end{align*} 
with the proper normalization constant $Z$.
Then the Seebeck coefficient is given by the mean  
\begin{align}\label{eq:Seebeck}
S &= -\avg{\omega} \; \beta\;,
\end{align}
and the Lorenz ratio
\begin{align}\label{eq:Lorenz:b}
L &= 
\avg{ \big(\Delta \omega\big)^{2}} \beta^{2}
\end{align}
by the variance of $\rho(\omega)$ and, consequently, the electronic part of the thermal conductance by
\begin{align}
\kappa_{e} &= \sigma  \avg{ \big(\Delta \omega\big)^{2}} \beta\;.
\end{align}

For low temperatures (degenerate electrons gas),  the variance of $\rho(\omega)$ is governed by the factor $|f'(\omega)|$, which is strongly localized in the interval
 ${\cal I}_{T}:=[-T,T]$,   in which ${\cal T}(\omega)$ is commonly assumed to be slowly varying. Then we obtain 
\begin{align}\label{eq:aux1}
\beta^{2} \avg{ \big(
\Delta \omega
\big)^{2}} &= \frac{\pi^{2}}{3}\;,
\end{align}
which corresponds to the Wiedemann-Franz law. We also realize that deviations are to be expected in the case when $T(\omega)$ is strongly varying  in  the interval  $\omega\in {\cal I}_{T}$, which is the case e.g. for the 2D tight-binding density of states, that has a van Hove singularity at $\omega=0$. We see that the electrical conductivity according to \eq{Conductance} is always positive. The same holds true, based on \eq{eq:Lorenz:b}, for the Lorenz ratio. Then, due to \eq{Lorenz_ratio} also the electronic part of the heat conductivity is positive. \eq{J_Phi_Delta_T} then guarantees that, given a bias voltage and no temperature difference, the electrical current flows from the side of higher to the lower potential energy. In the opposite case,
of zero bias voltage but finite $\Delta T$, inserting \eq{J_Phi_Delta_T} into yields \eq{JQ_Phi_Dela_T} 
\begin{align*}
J_{en} &= \big(T S^{2} \sigma + \kappa_e\big)\Delta T\;,
\end{align*}
and, since the bracketed expression is positive, the energy current flows from the side of higher temperature  to that of lower temperature. The same is valid for the heat current, which is given  by $\kappa_{e} \Delta T$. A sign change can, however, occur in the Seebeck coefficient, which according to \eq{eq:Seebeck} is a measure of the asymmetry of the transmission function. In the present model ${\cal T}(\omega)$ is symmetric about $\omega=0$ for $\varepsilon_{c}=0$.
Qualitatively, ${\cal T}(\omega)$ is shifted to positive frequencies by positive values of $\varepsilon_{c}$ and vice versa. Moreover, by definition, $S$ is an anti-symmetric function in $\varepsilon_{c}$.

The key element of the thermoelectric figure of merit is the Seebeck coefficient in \eq{eq:Seebeck},  that 
increases with the asymmetry of the probability distribution function $\rho(\omega)\propto |f'(\omega)| {\cal T}(\omega)$. The first factor is a narrow symmetric peak with exponential tails $e^{-\beta |\omega]}$. The asymmetry enters via the transmission function. If the latter is smooth, the mean $\avg{\omega}$ is restricted by $|f'|$ and the  asymmetry is at most $|k_{B}T|$, resulting in $S\lessapprox 1$. In order to have a figure of merit greater than 1, the transmission function has to overcome the peak $|f'(\omega)|$. This can be achieved by a transmission function that is zero up to some energy $\omega_{1}> T$ and has a sharp onset at $\omega_{1}$. In the  extreme case that ${\cal T}$ is a box function in the interval $(\omega_{1},\omega_{2})$, with $\omega_{2}\gg \omega_{1}\gg T$ then  
\begin{align*}
\rho(\omega) &\approx \frac{1}{Z}\;\Theta(\omega - \omega_{1}) e^{-\beta \omega} \;.
\end{align*}
In this case $S = 1+ \beta \omega_{1}$. This is actually the case in the high ZT materials reported  in \cite{ou.yu.16}.

\begin{center}
\begin{figure}[t!]
\subfigure[$\quad$2D tight-binding DOS.]{
\label{fig:with_2D_Dos}
\begin{minipage}[b]{0.44\textwidth}
\centering \includegraphics[width=1\textwidth]{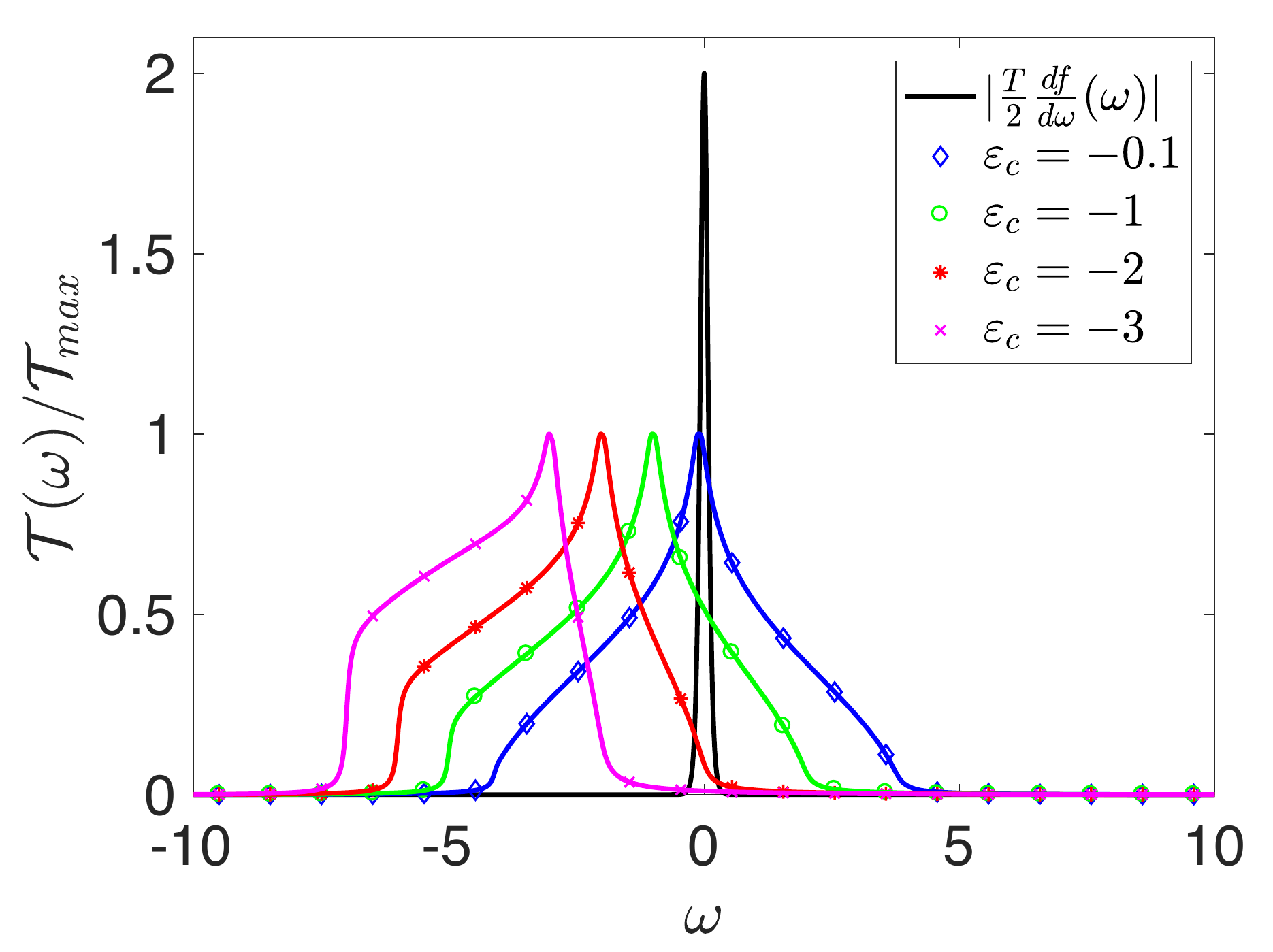}
\end{minipage}}\\
\hspace{0.025\textwidth}
\subfigure[$\quad$DOS of a Bethe lattice.]{
\label{fig:with_Bethe_Dos}
\begin{minipage}[b]{0.44\textwidth}
\centering \includegraphics[width=1\textwidth]{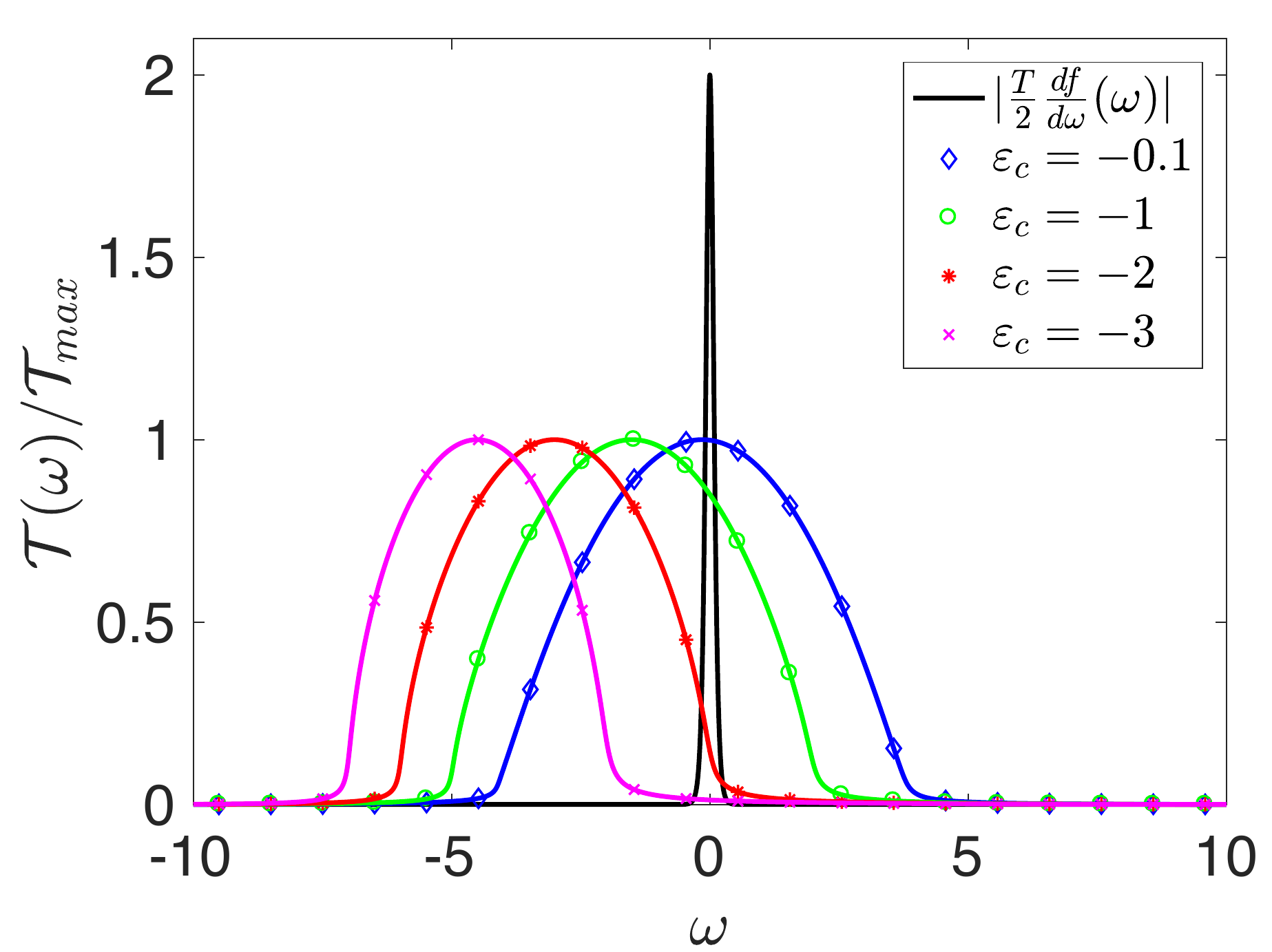}
\end{minipage}
}
\\
\caption{(Color online) Transmission function ${\cal T}(\omega)$ for $U=0$ and different values of $\varepsilon_{c}$ contrasted with $|f'(\omega)|$. Subfigure (a) corresponds to the results obtained by a 2D 
tight-binding DOS, while the results in subfigure (b) are produced by the Bethe lattice DOS. Both models have the same bandwidth. The asymmetry within the transport window, which determines the Seebeck coefficient, is most pronounced when the 
 'van Hove singularity' or   the edge of the transmission function passes the edge of the transport window, defined by $|f'(\omega)|$.}
\label{fig:Non_interacting_transmission}
\end{figure}
\end{center}

For non-interacting electrons in the model under consideration, the transmission function is non-zero in the interval ${\cal I}_{0}:=[-2-\gamma(0)+\varepsilon_{c},2+\gamma(0)+\varepsilon_{c}]$ with Lorentzian edges, whose locations  are shifted with  $\varepsilon_{c}$ (See Fig. \ref{fig:Non_interacting_transmission}). Due to the van Hove singularity the transport function ${\cal T}(\omega)$ has sharper peaks compared to the transport function  for a system without van Hove singularity, e.g. with a Bethe lattice DOS (compare Figs. \ref{fig:with_2D_Dos} and \ref{fig:with_Bethe_Dos}). It leads to the different behavior of $S$ when the maximum passes the peak of $|f'(\omega)|$. Moving the edges  of the transmission function via $\varepsilon_{c}$ across the peak of $|f'(\omega)|$ will increase $S$ but only until the edge is out of the energy window $\pm T$, then it will decease again and consequently, $S$ has a pronounced peak but the height will be $S\lessapprox 1$ for reasons outlined  before.

\section{Results}\label{Results} 

The emphases of the present paper is on the influence of strong electronic correlations on the thermoelectric properties, which will be studied in the framework  of the device displayed in Fig.~\ref{schematicp}, consisting of a correlated layer, with local Hubbard interaction $U$ and on-site energy $\varepsilon_c$, coupled to two metallic leads. Of particular interest is the impact of the various model parameters. We will first study the linear response  behavior of the system, for which only equilibrium calculations are necessary, before we move on to full nonequilibrium studies for strong temperature or  potential differences between the leads.

In all our calculations the hopping between neighboring sites inside the correlated layer is considered as energy unit ($t_c=1$), while the hopping 
between neighboring sites of the leads is always chosen as $t_l=t_r=2$.  As far as
the other parameters are concerned, we have distinguished between the linear response and the full non-equilibrium calculation. In the first case
we have chosen for the hopping into and out of the central region $v_l=v_r=0.25$ and for the temperature in the leads $T_l=T_r=0.025$, while  in the second case
we have used $v_l=v_r=1$ and $T_l=0.9$ and $T_r=0.7$, respectively,
in order to enhance the effect of the temperature  difference.

\subsection{Linear response}\label{Results_linear_response}

We start out with the results for the linear response regime. 
The temperature in the left and right lead is $T=0.025$ and the hopping  in and out of the central layer is $v_{l}=v_{r}=0.25$. Due to the particle-hole symmetry for $\varepsilon_{c}=0$, the electrical and thermal conductance are even functions of the onsite energy $\varepsilon_{c}$, while  the Seebeck coefficient is an odd function of $\varepsilon_{c}$. Therefore, we restrict our results to $\varepsilon_c \leq 0$, which corresponds  to fillings $n\geq 1$ in the central layer.

\subsubsection{Non-interacting particles}

\begin{figure}[t!]
\centerline{\includegraphics[width=0.85\columnwidth]
{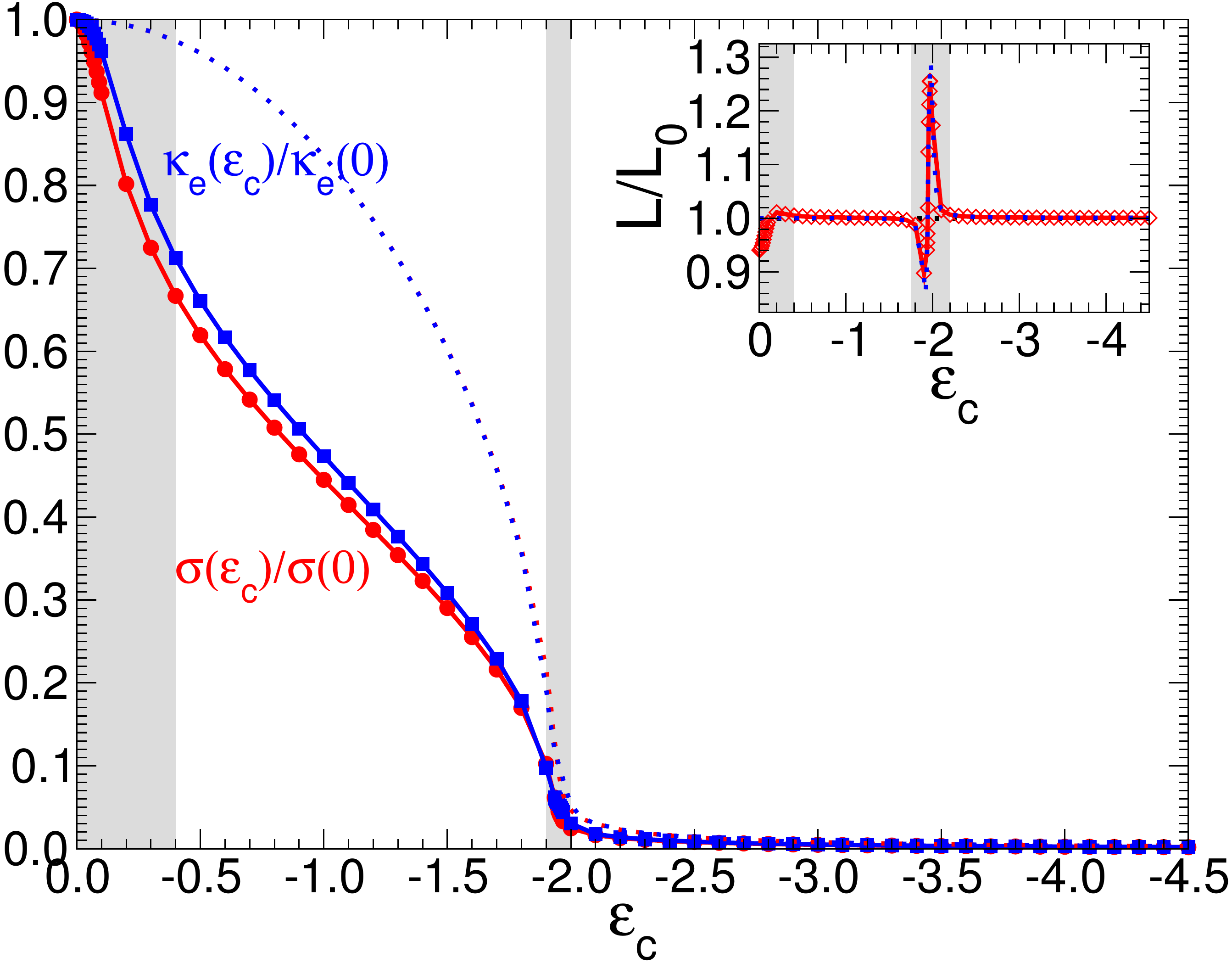}
}
\caption{Electrical and  thermal conductance  ($\sigma (\varepsilon_c)/\sigma (0)$ and $\kappa_e(\varepsilon_c)/\kappa_e(0)$) as a function of onsite energy 
$\varepsilon_c$ for the  non-interacting system. The other parameters are  
$t_{c}=1, t_{l/r}=2, v_{l/r}=0.25, T=0.025$. Dotted lines correspond to the results obtained for the Bethe lattice density of states with the same band-width (see main text). 
In case of the Bethe lattice, the two type of conductance differ mainly in the vicinity of $\varepsilon_{c}=-2$, therefore it is hard to distinguish them from each other.
Inset: Lorenz number $L/L_{0}$  (see Eq.~\eqref{Lorenz_ratio}).  Deviations from the Wiedemann-Franz law occur only at $\varepsilon_{0}\to 0$, due to a van Hove singularity and at $|\varepsilon_{0}|= 2$, 
due to a sharp edge in the transmission function ${\cal T}(\omega)$.}
\label{fig:Non_interacting}
\end{figure}

In order to single out the impact of correlations on the thermoelectric properties, we will first study the present model  for $U=0$. In this case the self-energy $\Sigma(\omega)$ vanishes and all required quantities for the integral Eq. \eqref{I_m} are given analytically and we can easily  determine the integral by numerical means. The results for the electrical and thermal conductance are presented in Fig.~\ref{fig:Non_interacting}. We find that both have a maximum  at $\varepsilon_c=0$ (half-filling),  decrease  monotonically  with increasing 
$|\varepsilon_{c}|$, and vanish rapidly beyond $|\varepsilon_{c}|=2$ . As pointed out before, both quantities only depend on $|\varepsilon_{c}|$.  

Furthermore, we have examined the validity of the Wiedemann-Franz law (see Eq.~\eqref{WFL}). In the inset of  Fig.~\ref{fig:Non_interacting} we present $L/L_0$ as a function of the onsite energy. We observe deviations from the  Wiedemann-Franz law only for values 
$|\varepsilon_{c}|\approx 0.1$ and $|\varepsilon_{c}|\approx 2.0$.
The first deviation is due to a pronounced peak in the transmission function ${\cal T}(\omega)$ that stems from the van Hove singularity of the 2D tight-binding density of states (see \eq{eq:2Ddos}) and the second deviation is due to a sharp edge in ${\cal T}(\omega)$ that stems from the finite width of 
$\gamma_{l/r}(\omega,t_{\alpha}\varepsilon)$. The latter is given by the imaginary part of the Green's function of the leads. In the wide band limit it is proportional to
\begin{align*}
\tilde\gamma_{l/r}(\omega,\varepsilon) &\propto \sqrt{
1- \big(\frac{\varepsilon}{2}\big)^{2}}\;.
\end{align*}

\begin{figure}[t!]
\centerline{\includegraphics[width=0.85\columnwidth]
{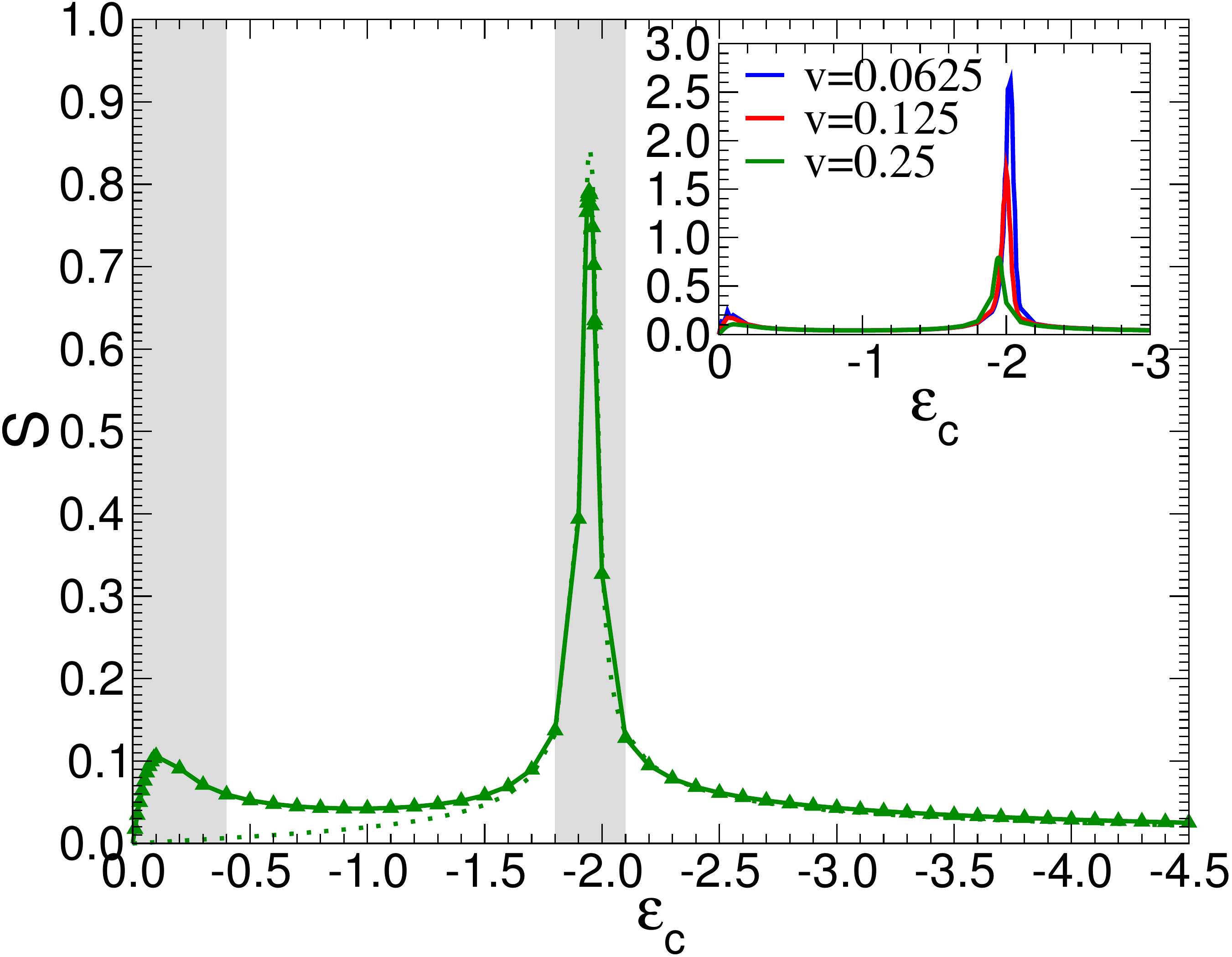}
}
\caption{Seebeck coefficient for the non-interacting system. Parameters as in \fig{fig:Non_interacting}. The dotted lines correspond to the results obtained by the Bethe lattice density of states, where the initial   hump is missing. The inset depicts  the Seebeck-coefficient versus  $\varepsilon_{c}$ for different values
of the coupling to the leads $v=v_{l/r}$. Obviously, the peaks increase with decreasing $v$.}
\label{fig:Non_interacting:Seebeck}
\end{figure}

On the other hand, the Seebeck coefficient $S(\varepsilon_c)$, which is depicted in 
\fig{fig:Non_interacting:Seebeck},
is a non-monotonic function and has two maxima: the first maximum is close 
to half-filling $\varepsilon_c \simeq -0.1$, and the  second at $\varepsilon_c \simeq -2$.  As pointed out in section \ref{Thermoelectric_properties},
maxima in $S$ occur, when sharp edges or maxima in the transmission function
are moved through the energy window ${\cal I}_{T}$ defined by $|f'|$.
The situation is illustrated in \fig{fig:Non_interacting_transmission}, where $|f'(\omega)|$ and the transmission function ${\cal T}(\omega)$ are shown for different values of $\varepsilon_{c}$. Obviously, the first maximum  is due the 
van Hove singularity, which for  $\varepsilon_{c}\simeq -0.1$ crosses the boundary of ${\cal I}_{T}$.  The  second maximum occurs at $\varepsilon_{c}\simeq -2$, when the edge of the transmission function crosses the boundary. The second maximum of the Seebeck coefficient is higher, because the edge of ${\cal T}$ is more pronounced than the van Hove singularity, resulting in a greater mean $\avg{\omega}$.

In order to underpin the impact of the van Hove singularity,  we have also performed calculations for the density of states $\rho_{B}(\epsilon)$ of the Bethe lattice, which has no singularity at $\omega=0$. 
$\rho_{B}(\epsilon)$ is chosen such that it has the same band width as the original density of state. We  find that both low-frequency features, the maximum of the Seebeck coefficient and the discrepancy 
from  the Wiedemann-Franz law, disappear.
The features at $\varepsilon_{c}=-2$ survive as they are due to band edge effects.

Finally, still for the non-interacting model, we also present the dependence of the the Seebeck coefficient on the hybridization $v_l=v_r=v$ in the inset of Fig.~\ref{fig:Non_interacting:Seebeck}. 
Clearly, the Seebeck coefficient increases with decreasing hybridization $v$.

\begin{figure}[t!]
\centerline{\includegraphics[width=0.85\columnwidth]
{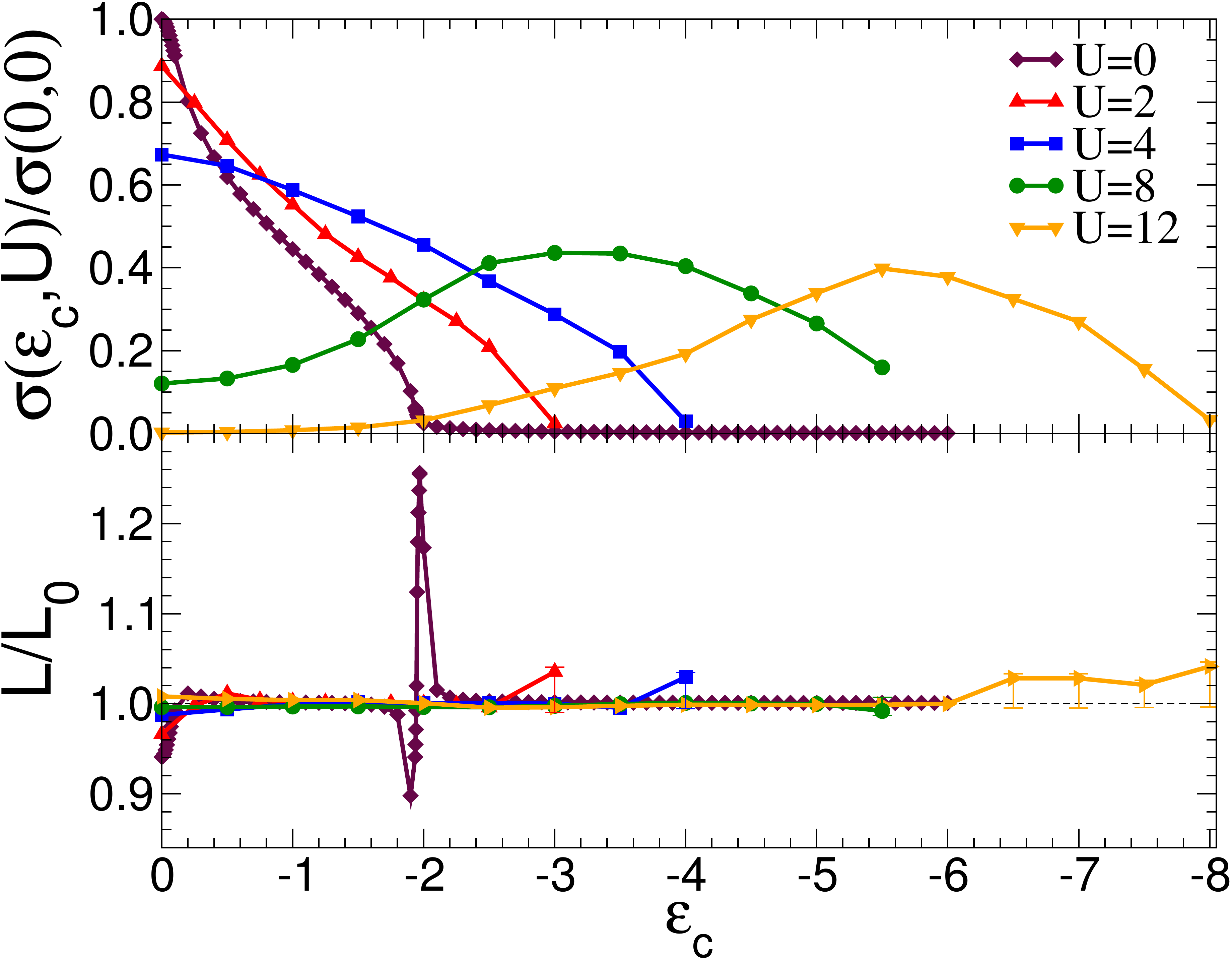}
}
\caption{Electrical conductance $\sigma$(upper panel) and Lorenz number (lower panel) as a function of  $\varepsilon_{c}$  for different values of $U$.
Calculations are performed with $N_b=6$. Other parameters see Fig.~\ref{fig:Non_interacting}.}
\label{fig:G&L_vs_eps}
\end{figure}

\subsubsection{Linear response for finite interaction}

\begin{figure}[t!]
\centerline{\includegraphics[width=0.85\columnwidth]
{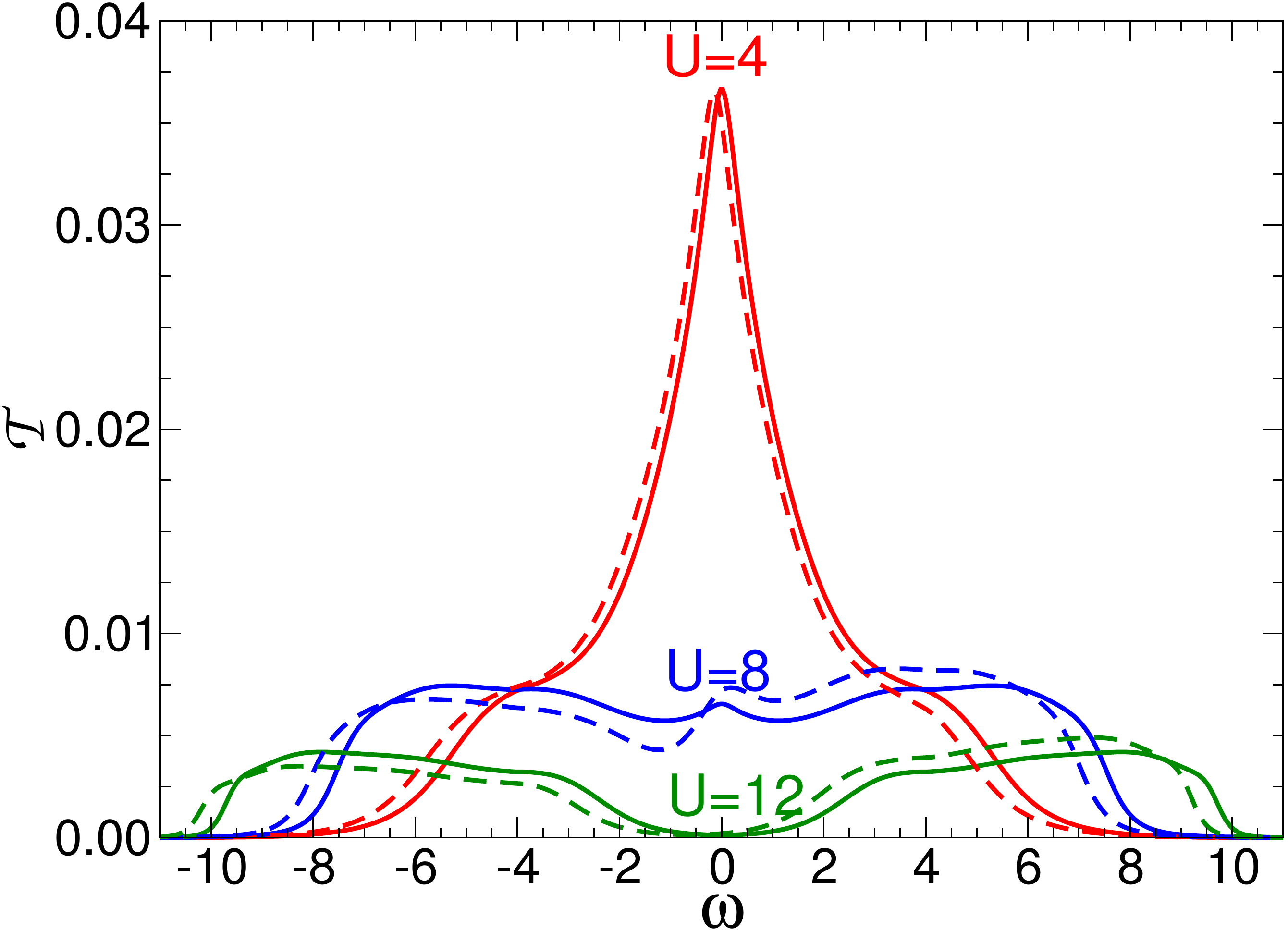}
}
\caption{Transmission function for different Hubbard parameters $U$. Solid line is for half-filling $\varepsilon_{c}=0$ and dashed line represents $\varepsilon_{c}= - 0.5$.
Other parameters as in  \fig{fig:Non_interacting}.}
\label{fig:transmission.u}
\end{figure}

Next we discuss the results for the interacting system. In the interacting case, we  first have to solve the self-consistent impurity problem and calculate the self-energy $\Sigma(\omega)$, before we can calculate thermoelectric properties introduced in Sec. \ref{Thermoelectric_properties}. In Fig.~\ref{fig:G&L_vs_eps} we show the electrical conductance $\sigma$ and the Lorenz number as a function of $\varepsilon_{c}$  for different values of the local interaction $U$. For small to moderate interactions ($U \lesssim 5.5$) we find, like in  the non-interacting case,  a decreasing  conductance  with increasing $\varepsilon_{c}$, which vanishes rapidly beyond $|\varepsilon_{c}|=2+U/2$. As a new aspect, the  maximum at $\varepsilon_{c}=0$ decreases with increasing $U$. For  strong interactions ($U \gtrsim 5.5$), on the other hand,  $\sigma$ behaves differently. A local minimum develops at $\varepsilon_{c}=0$ and a maximum at some finite value $\varepsilon^{*}_{c}$. Interestingly (not depicted in the plot), the filling corresponding to $\varepsilon^{*}_{c}$ is in all cases roughly the same $n \simeq 1.3$. Qualitatively, such non-monotonic behavior can be explained by the fact that there are two competing effects. On the one hand, like in the non-interacting case, with increasing $|\varepsilon_c|$ and correspondingly $n$ the conductance should decrease due to the decreasing overlap of  the spectral function of the central layer with the imaginary part of the leads Green's functions. On the other hand, close to half-filling ($\varepsilon_{c}=0$) correlation effects are most dominant and therefore due to backscattering, the conductance should be reduced or even suppressed. The latter is actually the case for  very strong interactions ($U \ge 12$), indicating that the isolated correlated interface would be in  the Mott insulator phase, but the system always remains metallic when the correlated layer is coupled to the leads. In the limit of strong interaction the conductance is strongly suppressed, but it never becomes equal to zero. The Mott transition is nicely visible in the transmission function, which is shown in \fig{fig:transmission.u}, where a gap is found at $U=12$.\cite{Gap} As far as the Lorenz number is concerned, which is depicted in the lower half of Fig.\ref{fig:G&L_vs_eps},  the deviation from the Wiedemann-Franz law, that we found in the non-interacting case at $\varepsilon_{c}\approx 0$, decreases with increasing interaction strength, because the van Hove singularity is smeared out by the interaction. 

\begin{figure}[t!]
\centerline{\includegraphics[width=0.85\columnwidth]
{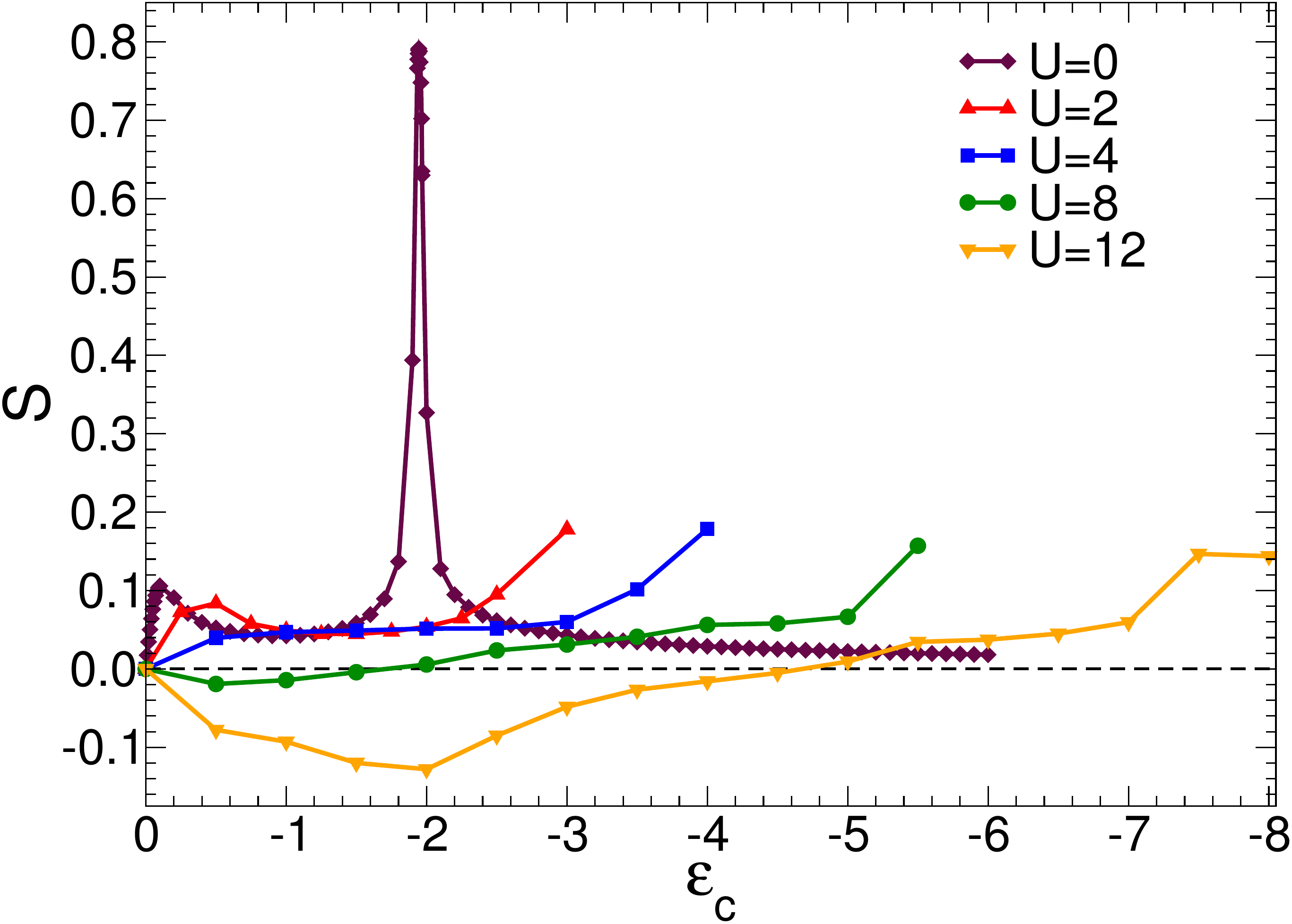}  
}
\caption{Seebeck coefficient $S$ as a function of $\varepsilon_{c}$ for different values of $U$. (Other parameters as in  \fig{fig:Non_interacting}).The hump close to half filling is due to the van Hove 
singularity at $\omega=0$ of the 2D tight-binding density of the state and the peak close to $n=1.6$ is due to the sharp drop of the transmission function ${\cal T}(\omega)$. 
}
\label{fig:S_vs_eps_v}
\end{figure}

\begin{figure}[t!]
\centerline{\includegraphics[width=0.85\columnwidth]
{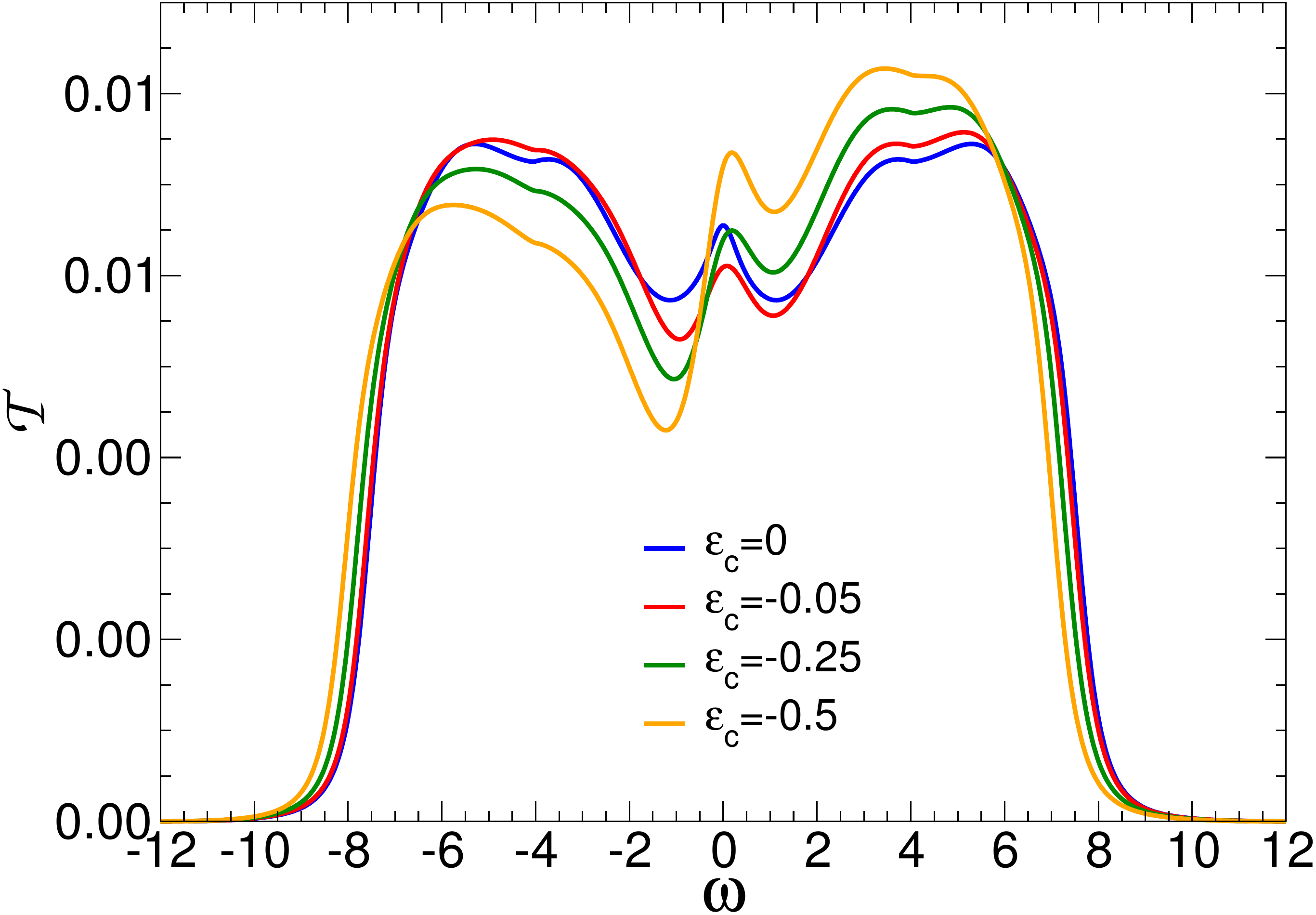}
}
\caption{Transmission function for  $U=8$ and different values of $\varepsilon_{c}$. Other parameters as in  \fig{fig:Non_interacting}.}
\label{fig:transmission.eps}
\end{figure}

As a final point on linear response, we turn to the  Seebeck coefficient (see Fig.~\ref{fig:S_vs_eps_v}). We find that for $U \lesssim 5.5 $ the Seebeck coefficient is positive, as in the non-interacting case. With increasing interaction the maxima are shifted to higher $\varepsilon_{c}$-values and are suppressed. Along the same line of reasoning we gave in the non-interacing case, this behavior can easily be explained by the smearing of the transmission function due to interactions. However, a qualitatively new aspect comes into play for strong interactions ($U \gtrsim 5.5$). 
Now we find that for small onsite energies $\varepsilon_{c}^{\star} \le \varepsilon_{c}\le 0$ the Seebeck coefficient becomes negative, before it turns positive again for some 
value $\varepsilon_{c}^{\star}$. This is in particular the case for $U=8$ and $U=12$ in 
\fig{fig:S_vs_eps_v}, while $U=4$ still exhibits a purely positive Seebeck coefficient.
To understand the qualitative difference in the Seebeck coefficient corresponding to these $U$ values, we consider the corresponding transmission functions in \fig{fig:transmission.u}.
We observe that the central peak (van Hove singularity) in the transmission function, which is still predominant for $U=4$, decreases  with increasing $U$ and the lower and upper Hubbard band grow, until eventually the Mott-Hubbard gap opens up 
(See Fig.~\ref{fig:transmission.u}). At the same time, for $U \gtrsim 5.5$, spectral weight is transferred from the lower to the upper Hubbard band, when $\varepsilon_c<0$. Consequently, based on Eq. \eqref{omegam}, the mean $\avg{\omega}$ moves to positive energies, resulting in a negative Seebeck coefficient, absolute value of which increases with increasing $U$. 
The detailed dependence of the transmission function on $\varepsilon_{c}$ for the case $U=8$ is given in \fig{fig:transmission.eps}. It is obvious, why  the Seebeck coefficient $S=-\beta \avg{\omega}$ is increasingly negative for increasing $\varepsilon_{c}$. In addition to the transfer of spectral weight from the lower to the upper Hubbard band, the entire transmission function shifts to the left with increasing $\varepsilon_{c}$, like in the non-interacting case. This shift eventually, for $\varepsilon_{c}< \varepsilon_{c}^{\star}<0$ results in a positive Seebeck coefficient.
As we can clearly see in \fig{fig:transmission.eps} for the $U=8$ case, the edge close to
$\omega=0$ becomes increasingly pronounced with increasing $\varepsilon_{c}$-values.
At the same time the Seebeck coefficient increaeses. 
This is in nice agreement with our general discussion that the Seebeck coefficient is large if a sharp edge or peak is shifted (by changing some parameter) across the energy window, defined by $|f'|$.
It is particularly interesting to note that the absolute height of the first peak in the Seebeck 
coefficient is greater for $U=12$ than in the non-interacting case. According to \eq{Figure_of_merit} the  thermoelectric figure of merit depends quadratically on $S$, therefore we find that strong correlations would increase  the figure of merit, taking into account that the Lorenz number even slightly supports this effect.

The behavior of $ZT$ is completely determined by the behavior of the Seebeck coefficient $S$ and the Lorentz number $L$. Even more, based on the sign of the Seebeck coefficient one can determine whether the conductance is electron like or hole like, which is impossible to do based on the figure of merit. Therefore we prefer to present the results for the Seebeck coefficient $S$ and the Lorentz number $L$ separately. 

\begin{figure}[t!]
\centerline{\includegraphics[width=0.85\columnwidth]
{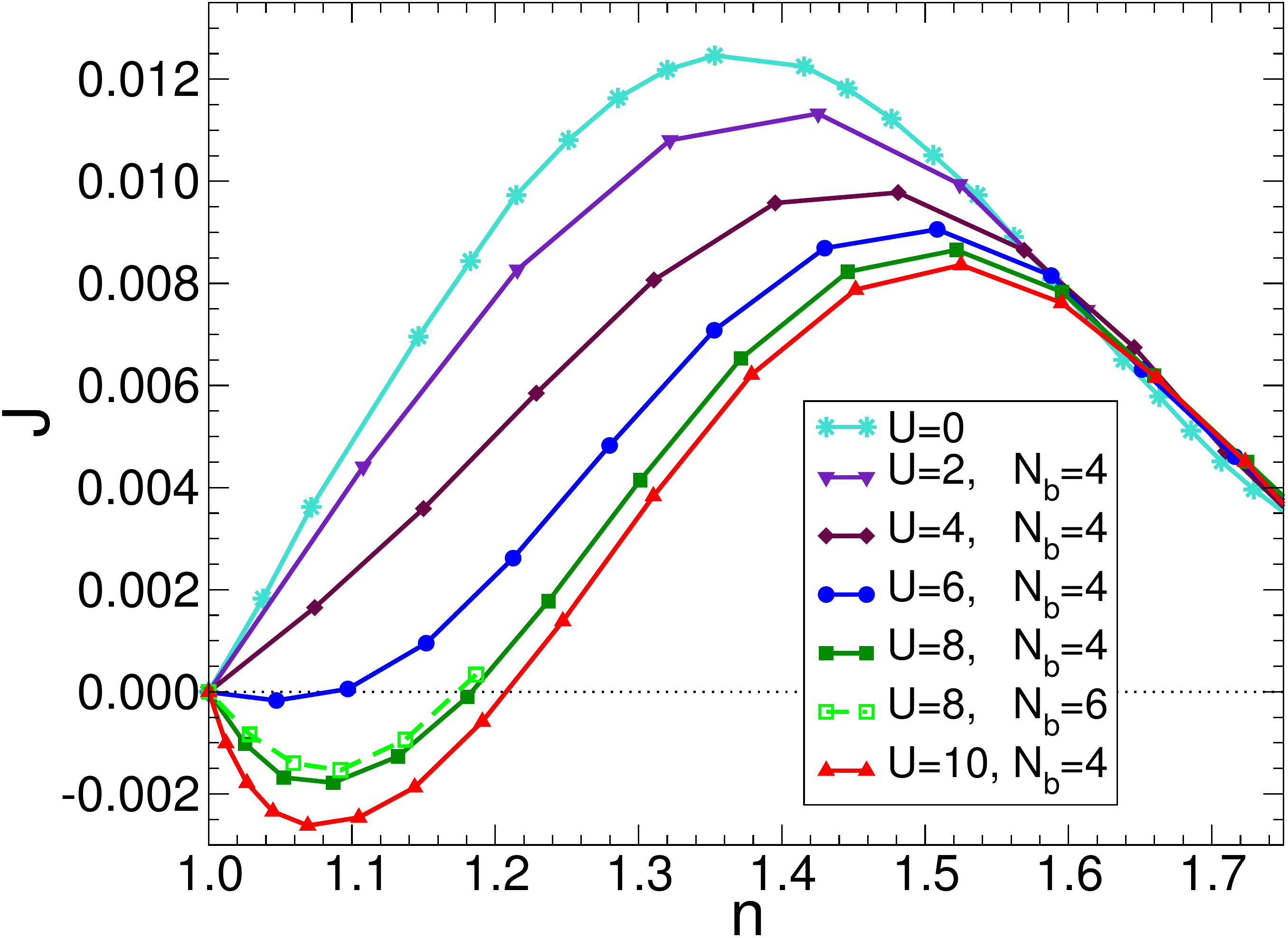}    }
\caption{Current density $J$ as a function of filling $n$ for  $T_l=0.9$, $T_r=0.7$ and different values of $U$. The other parameters are as before $t_l=t_r=2$, $v_l=v_r=1$, and ($\Phi=0$).
All calculations are performed with  $N_b=4$, except for $U=8$ we have also used 
$N_{b}=6$ to test convergence.}
\label{fig:Current_vs_N_Phi0}
\end{figure}
\begin{figure}[t!]
\centerline{\includegraphics[width=0.85\columnwidth]
{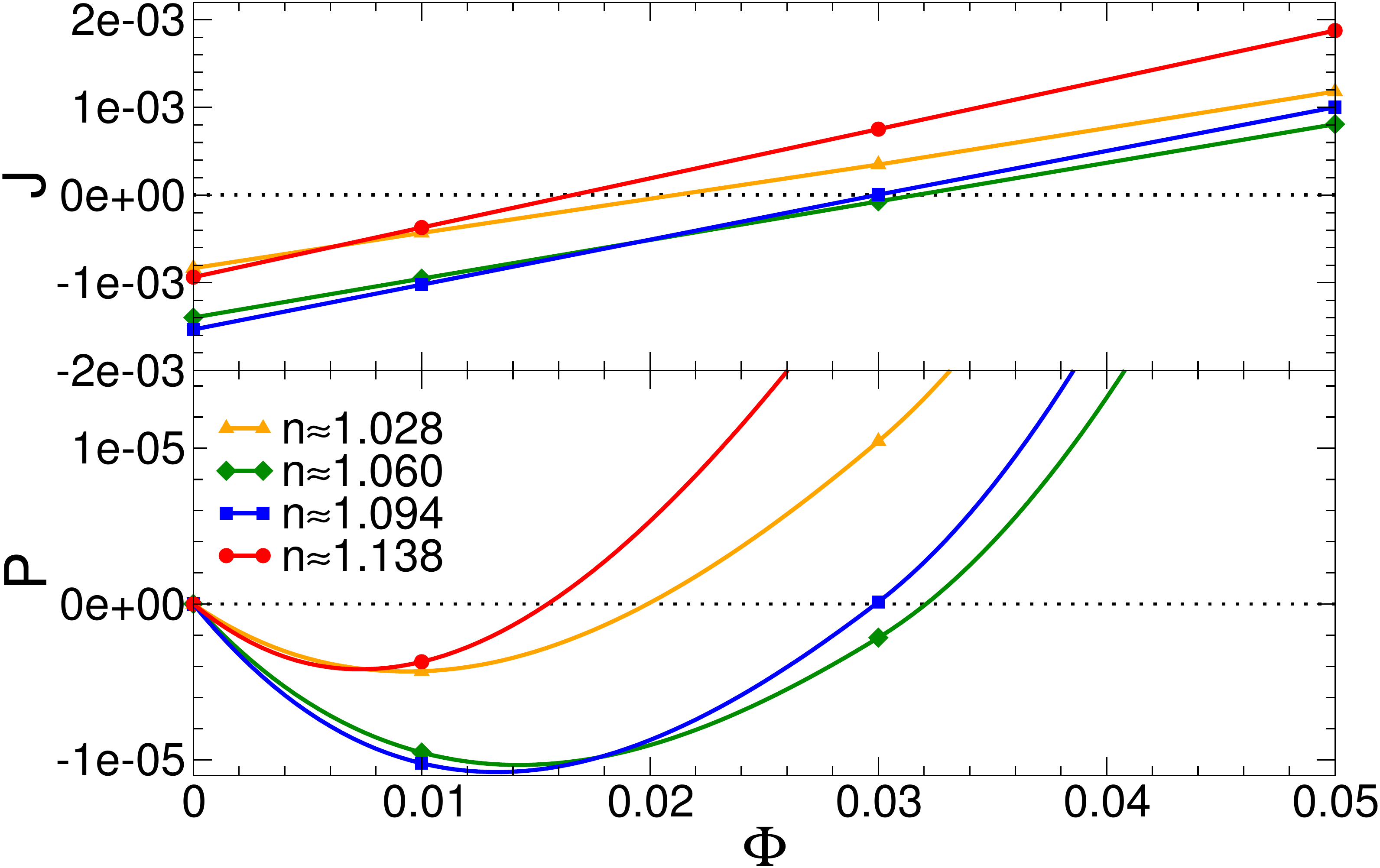}
}
\caption{Current density $J$ (upper panel) and power $P$ (lower panel) as a function of the bias voltage $\Phi$ for ${U=8}$ and different filling $n\simeq 1.028,\,1.060,\,1.094,\,1.138$. Other parameters are 
the same as in Fig.~\ref{fig:Current_vs_N_Phi0}. Calculations are performed with ${N_b=6}$. The lines in the upper panel are obtained by a straight line fit and the solid lines in the lower panel then follow 
from $P=J \Phi$.}
\label{fig:Power_vs_Phi_U8}
\end{figure}

\subsection{Finite temperature difference}\label{Results_temperature_gradient}

Following the discussion of linear response results, we investigate the effect of a finite temperature difference between the leads. 
We start out with zero  bias voltage  ($\Phi=0$), where the  nonequilibrium situation is driven merely by  the temperature difference (${\Delta T>0}$). 
To emphasize the effect of $\Delta T$ on the the behavior of the system, in our calculations the large temperature difference. Therefore, the temperature values in the leads are $T_{l}=0.9$ and $T_{r}=0.7$, respectively. It should be stressed, that the average temperature is much higher (non-physical) than the one we used in the linear response calculations, which was $T=0.025$. To obtain more smooth results  we consider the hopping into and out of the central region is now $v_{l}=v_{r}=1$ instead of $0.25$, which we used in the linear regime. In Fig.~\ref{fig:Current_vs_N_Phi0} we present the current density $J$ as a function of the filling  $n$ for different Hubbard interactions $U$. Due to particle-hole symmetry  ${J(n)=-J(2-n)}$. 

Again, in the non-interacting case, $U=0$, the self-energy is zero and all expressions are analytically available. In this case, the current is positive above the half-filling ($n>1$) and negative in the opposite case. We find that the current has a bell shape $n$-dependence, with a maximum at $n\simeq 1.375$, which corresponds to an onsite energy $\varepsilon_{c}\simeq -1.3$. The direction of the current can be explained by the two  factors $\Delta f := f_{l}(\omega)-f_{r}(\omega)$, which is an antisymmetric function about $\omega=\mu$ (positive above and negative below $\mu$) and ${\cal T}(\omega)$ in \eq{Current}. Above half filling ($n>1$), the transmission function ${\cal T}(\omega)$ has negative slope at the chemical potential (see e.g. \fig{fig:Non_interacting} for $\varepsilon_{c}<0$) and therefore the integral $\int \Delta f(\omega) {\cal T}(\omega) d\omega$ is dominated by the negative part of the integrand. Or in more physical terms: electrons move  from the cold right lead to the warm left lead, which can also be described by holes moving from the warm left lead to the cold right lead and  $J>0$. Below  half-filling, the opposite is the case. The slope of ${\cal T}(\omega)$ at the chemical is positive and, therefore, the current integral is dominated by the positive parts, which can be described by electrons moving from the warm left lead to the cold right lead and  $J<0$.

Next we turn to the interacting case. For $U=8$, we have first tested the convergence of our numerical scheme with respect to the number of bath sites in our impurity solver. In \fig{fig:Current_vs_N_Phi0} the corresponding results are depicted for $N_{b}=4$ and $N_{b}=6$. We find that the resulting curves can hardly be distinguished. Therefore, we restrict all further results to  $N_{b}=4$.

For the weak to intermediate interactions ($U \lesssim 5.5$), as might be expected, the behavior of the current $J$ as a function of the onsite energy $\varepsilon_{c}$ is qualitatively similar to the non-interacting case.  

The impact of $U$ is a shift of the current maximum  to higher $\varepsilon_{c}$-values, or rather higher fillings, and the value of the maximum decreases slightly. This behavior can qualitatively be explained by our previous findings in the linear response case. Accordingly it the expression  for the current, driven by the temperature difference, is $J \propto \sigma S$. Taking into account the much higher average temperature in the present case, the bell shaped structure, the shift of the maximum to higher $\varepsilon_{c}$ values, and the reduction of the maximum, can be inferred from \fig{fig:G&L_vs_eps} and \fig{fig:S_vs_eps_v}. The linear response expression would predict , due to the features of $S$, that for stronger interaction the current is  negative for $\varepsilon_{c}$ below a certain threshold $\varepsilon_{c}^{\star}$ and positive above. This is indeed the result, found in the full nonequilibrium many-body calculation, depicted in \fig{fig:Current_vs_N_Phi0}. We obtain also that  for strong coupling, the absolute value of the current increases with increasing interaction strength. In the vicinity of half-filling ($1<n\lesssim 1.05$) the current is larger as compared to the non-interacting system with the same filling. With further increase of the filling $n$, correlation effects become less relevant, therefore the current decreases and changes sign. The change of sign of the current takes place for the higher fillings with increasing the interaction. Finally, for even higher fillings, close to the full filling ($n \gtrsim 1.6$), interaction plays even less important role in the behavior of the system and correspondingly the current is nearly independent from it.

Finally, we consider the case, when in addition to the finite temperature difference mentioned above also a finite bias voltage $\Phi$ is applied, and investigate the current-voltage characteristics.
In this case \eq{Current} is not valid anymore, therefore we used the more general Meir-Wingreen formula\cite{ha.ja, me.wi.92, jauh, re.zh.12, ti.do.15}. In the upper panel of \fig{fig:Power_vs_Phi_U8} the current density is plotted versus applied voltage for different fillings, which are tuned by $\varepsilon_{c}$. The fillings are chosen such, that the current without bias voltage is negative. We find a linear increase, which agrees with the linear response result, obtained by \eq{J_Phi_Delta_T}.

From the current-voltage curve, we can directly compute the power $P=J\cdot\Phi$. Obviously, it is negative for ${0<\Phi<\Phi_0(\varepsilon_c,U)}$ (see Fig.~\ref{fig:Power_vs_Phi_U8}[lower panel]), which means that energy conversion from the system takes place. The maximum efficiency of the device is reached for that bias voltage for which the absolute value of the power $P$ reaches its maximum. One can readily see that this is the case for  ${\Phi \simeq \frac{1}{2}\Phi_0(\varepsilon_c, U)}$. The corresponding power is 
$P \simeq \frac{1}{2}J_0(\varepsilon_c,U)\Phi_0(\varepsilon_c, U)$. Our calculations have shown that for $U \gtrsim 5.5$ the maximum value of $|J_0(\varepsilon_c,U)|$ as well as $\Phi_0(\varepsilon_c, U)$  increases with increasing  interaction and the optimal filling varies between $1.04<n<1.08$. So, similar to what we have found in the linear response case, the efficiency of the thermoelectric device can be increased in  strongly correlated materials close to half-filling.

\section{Conclusions}\label{Conclusions}

In this work we have investigated  linear response and steady state properties of a device, consisting of a correlated mono-layer attached to two metallic leads. In the full many-body nonequilibrium calculation the nonequilibrium properties were driven by a temperature difference in the leads and a bias voltage. The key quantity, as far as the thermoelectric efficiency is concerned, is the thermoelectric
figure of merit, which is proportional to the square of the Seebeck coefficient and inverse proportional to the Lorenz number. As the latter is very well described by the Wiedemann-Franz law, the efficiency is governed solely by the Seebeck coefficient. We have given some general arguments that a large Seebeck number requires sharp structures in the transmission function, which indeed occurred for strong interaction strength $U$. We have shown that the efficiency of the thermoelectric device can furthermore be increased by increasing the overall temperature and/or decreasing the coupling strength between device and leads. But more importantly, we have demonstrated, both in the linear response regime and beyond, that  the efficiency of the device can be increased by strong correlations close to half-filling.

\appendix
\section{Generalisation of the Meir-Wingreen formula for a layered system}\label{Appendix:Meir-Wingreen}

Here we generalize the  Meir-Wingreen formula to the case of a layered system and derive equation \eqref{Current}. 

We start out from the expression\cite{okam.07, Note_appendix}  for the current from layer $z$ to layer $z+1$, for which the hopping matrix element in the second term of equation \eqref{Hamiltonian} is $t_{z,z+1}$
\begin{eqnarray}
I&=&\frac{e t_{z,z+1}}{\hbar}\int \frac{(dk_{||})^2d\omega}{(2\pi)^3}\left\{G^K_{z+1,z}-G^K_{z,z+1}\right\} \notag\\
&=& \frac{2e t_{z,z+1}}{\hbar}\int \frac{(dk_{||})^2d\omega}{(2\pi)^3} \mathrm{Re}G^K_{z+1,z} \label{app:MW}\\
&=& -\frac{2e t_{z,z+1}}{\hbar}\int \frac{(dk_{||})^2d\omega}{(2\pi)^3} \mathrm{Re}G^K_{z,z+1} \;.\notag
\end{eqnarray}

We have exploited  the fact that the Keldish Green's function ${\bf G}^K$ is antihermitian ($G^K_{z+1,z}=-(G^K_{z,z+1})^*$). Due to the  conserved  current  the results are  independent of  $z$. Then we find for 
the current $I_{r/l}$, that flows from the right/ left lead into the central layer,
the expression
\begin{align}
I_{r/l} &=  \frac{2e v_{r/l}}{\hbar}\int \frac{(dk_{||})^2d\omega}{(2\pi)^3} \mathrm{Re}G^K_{\pm1,0} \;.\label{eq:MW:lr}
\end{align}
Next we will express the off-diagonal elements $G^{K}_{\pm1,0}$ by  diagonal elements of suitable Green's functions. 

Since the first index of the Green's function ($z=\pm 1$)  belongs to sites 
of the left or right lead where there is no Hubbard interaction,
 the equation of motion \cite{niu_equation_1999} simply yields
\begin{eqnarray*}
&&G^K_{\mp 1,0}=-v_{l/r} \left(G_{0,0}^R \;g^K_{l/r} + G_{0,0}^K \;g^A_{l/r}\right)\;,
\end{eqnarray*}
were we have used  that $g^{\gamma}_{\mp 1,\mp 1}=g^{\gamma}_{l/r}$
is the surface Green's function of the isolated left/ right lead.
Based on the general property of Green's functions $G^{A}=(G^{R})^{\dagger}$ and on definition in equation \eqref{eq:gamma} we find
\begin{widetext}
\begin{eqnarray*}
v_{l/r}~\mathrm{Re}G^K_{\mp 1,0}&=&-\frac{1}{2}v_{l/r}^2\left[\left(G_{c}^R-G_{c}^A\right)g^K_{l/r}-G^K_{c}\left(g^R_{l/r}-g^A_{l/r}\right)\right]
                =-\frac{1}{2}v_{l/r}^2\left[\left(G_{c}^R-G_{c}^A\right)2i(1-2f_{l/r})\mathrm{Im}g^R_{l/r} - G^K_{c}2i\mathrm{Im}g^R_{l/r}\right]\\
                &=&-\frac{i}{2}\left[2\left(G_{c}^R-G_{c}^A\right)f_{l/r}+ G^K_{c}-\left(G_{c}^R-G_{c}^A\right)\right]\gamma_{l/r}
                =-i\left[\left(G_{c}^R-G_{c}^A\right)f_{l/r}+ G^<_{c}\right]\gamma_{l/r} \, ,
\end{eqnarray*}
\end{widetext}
where we have introduced the lesser Green's function $G^{<}$, for which 
the general relation $G^{<} =G^K_{c}- (G_{c}^R-G_{c}^A)$ applies. 
Here $f_{l/r}$ is Fermi functions for the left and right lead, respectively. 
Inserting the above expressions for $G^{K}_{\mp1,0}$ into the Meir-Wingreen formula for the current $I_{l/r}$  (equation \eqref{eq:MW:lr}) we obtain
\begin{eqnarray*}
I_{l/r}&=&\frac{i e}{2\hbar}\int \frac{(dk_{||})^2d\omega}{(2\pi)^3}
\gamma_{l/r} 
\left[G_c^< + f_{l/r} \left(G_c^R -
G_c^A\right) \right] \, .
\end{eqnarray*}
Due to the current conservation we have $I=I_l=-I_r$ and for arbitrary $x$
we can also express the current as $I=xI_l-(1-x)I_r$.
Following the restricting assumption of Jauho\cite{jauh} that $\gamma_l =\lambda' \gamma_r$, we obtain
\begin{eqnarray*}
I&=&\frac{i e}{2\hbar}\int \frac{(dk_{||})^2d\omega}{(2\pi)^3} \gamma_r
\left[\left(\lambda' x- (1-x)\right) G_c^< \right. \\
&+&\left.\left(\lambda' x f_l - (1-x) f_r\right)\left(G_c^R -
G_c^A\right) \right] \, .
\end{eqnarray*}
Now if we fix the arbitrary parameter by $x=1/(1+\lambda')$, so that the term containing $G^{<}$ vanishes,  and taking into
account the relation $G_c^R - G_c^A =-2iA$,  we obtain 
\begin{eqnarray*}
I=\frac{e}{\hbar}\int \frac{(dk_{||})^2d\omega}{(2\pi)^3} \left(f_l -
f_r\right)
\frac{2\pi\gamma_l(\omega,{\vv  k})\gamma_r(\omega,{\vv 
k})}{\gamma_l(\omega,{\vv  k})+\gamma_r(\omega,{\vv  k})}A \, . \\
\end{eqnarray*}
In the units, used in this paper, this is nothing else but Eq. \eqref{Current} 
along  with equation \eqref{eq:Twk}, if the $k_{||}$-integration is replaced by the energy integration.

\begin{acknowledgments}

This work was supported by the Austrian Science Fund (FWF):  P26508, as well as SfB-ViCoM project F04103, and NaWi Graz. The calculations were partly performed on the D-Cluster Graz 
and on the VSC-3 cluster Vienna. 
\end{acknowledgments}


\begin{thebibliography}{67}%
\makeatletter
\providecommand \@ifxundefined [1]{%
 \@ifx{#1\undefined}
}%
\providecommand \@ifnum [1]{%
 \ifnum #1\expandafter \@firstoftwo
 \else \expandafter \@secondoftwo
 \fi
}%
\providecommand \@ifx [1]{%
 \ifx #1\expandafter \@firstoftwo
 \else \expandafter \@secondoftwo
 \fi
}%
\providecommand \natexlab [1]{#1}%
\providecommand \enquote  [1]{``#1''}%
\providecommand \bibnamefont  [1]{#1}%
\providecommand \bibfnamefont [1]{#1}%
\providecommand \citenamefont [1]{#1}%
\providecommand \href@noop [0]{\@secondoftwo}%
\providecommand \href [0]{\begingroup \@sanitize@url \@href}%
\providecommand \@href[1]{\@@startlink{#1}\@@href}%
\providecommand \@@href[1]{\endgroup#1\@@endlink}%
\providecommand \@sanitize@url [0]{\catcode `\\12\catcode `\$12\catcode
  `\&12\catcode `\#12\catcode `\^12\catcode `\_12\catcode `\%12\relax}%
\providecommand \@@startlink[1]{}%
\providecommand \@@endlink[0]{}%
\providecommand \url  [0]{\begingroup\@sanitize@url \@url }%
\providecommand \@url [1]{\endgroup\@href {#1}{\urlprefix }}%
\providecommand \urlprefix  [0]{URL }%
\providecommand \Eprint [0]{\href }%
\providecommand \doibase [0]{http://dx.doi.org/}%
\providecommand \selectlanguage [0]{\@gobble}%
\providecommand \bibinfo  [0]{\@secondoftwo}%
\providecommand \bibfield  [0]{\@secondoftwo}%
\providecommand \translation [1]{[#1]}%
\providecommand \BibitemOpen [0]{}%
\providecommand \bibitemStop [0]{}%
\providecommand \bibitemNoStop [0]{.\EOS\space}%
\providecommand \EOS [0]{\spacefactor3000\relax}%
\providecommand \BibitemShut  [1]{\csname bibitem#1\endcsname}%
\let\auto@bib@innerbib\@empty
\bibitem [{\citenamefont {Benenti}\ \emph {et~al.}(2013)\citenamefont
  {Benenti}, \citenamefont {Casati}, \citenamefont {Prosen},\ and\
  \citenamefont {Saito}}]{be.ca.13u}%
  \BibitemOpen
  \bibfield  {author} {\bibinfo {author} {\bibfnamefont {G.}~\bibnamefont
  {Benenti}}, \bibinfo {author} {\bibfnamefont {G.}~\bibnamefont {Casati}},
  \bibinfo {author} {\bibfnamefont {T.}~\bibnamefont {Prosen}}, \ and\ \bibinfo
  {author} {\bibfnamefont {K.}~\bibnamefont {Saito}},\ }\href
  {http://arxiv.org/abs/1311.4430} {\enquote {\bibinfo {title} {Fundamental
  aspects of steady state heat to work conversion},}\ } (\bibinfo {year}
  {2013}),\ \bibinfo {note} {arXiv:1311.4430}\BibitemShut {NoStop}%
\bibitem [{\citenamefont {Benenti}\ \emph {et~al.}(2017)\citenamefont
  {Benenti}, \citenamefont {Casati}, \citenamefont {Saito},\ and\ \citenamefont
  {Whitney}}]{be.ca.17}%
  \BibitemOpen
  \bibfield  {author} {\bibinfo {author} {\bibfnamefont {G.}~\bibnamefont
  {Benenti}}, \bibinfo {author} {\bibfnamefont {G.}~\bibnamefont {Casati}},
  \bibinfo {author} {\bibfnamefont {K.}~\bibnamefont {Saito}}, \ and\ \bibinfo
  {author} {\bibfnamefont {R.}~\bibnamefont {Whitney}},\ }\href {\doibase
  https://doi.org/10.1016/j.physrep.2017.05.008} {\bibfield  {journal}
  {\bibinfo  {journal} {Physics Reports}\ }\textbf {\bibinfo {volume} {694}},\
  \bibinfo {pages} {1 } (\bibinfo {year} {2017})}\BibitemShut {NoStop}%
\bibitem [{\citenamefont {Kirchner}, \citenamefont {Zamani},\ and\
  \citenamefont {Mu{\~{n}}oz}(2013)}]{ki.za.13}%
  \BibitemOpen
  \bibfield  {author} {\bibinfo {author} {\bibfnamefont {S.}~\bibnamefont
  {Kirchner}}, \bibinfo {author} {\bibfnamefont {F.}~\bibnamefont {Zamani}}, \
  and\ \bibinfo {author} {\bibfnamefont {E.}~\bibnamefont {Mu{\~{n}}oz}},\
  }\enquote {\bibinfo {title} {Nonlinear thermoelectric response of quantum
  dots: Renormalized dual fermions out of equilibrium},}\ in\ \href {\doibase
  10.1007/978-94-007-4984-9_10} {\emph {\bibinfo {booktitle} {New Materials for
  Thermoelectric Applications: Theory and Experiment}}},\ \bibinfo {editor}
  {edited by\ \bibinfo {editor} {\bibfnamefont {V.}~\bibnamefont {Zlatic}}\
  and\ \bibinfo {editor} {\bibfnamefont {A.}~\bibnamefont {Hewson}}}\ (\bibinfo
   {publisher} {Springer Netherlands},\ \bibinfo {address} {Dordrecht},\
  \bibinfo {year} {2013})\ pp.\ \bibinfo {pages} {129--168}\BibitemShut
  {NoStop}%
\bibitem [{\citenamefont {Harman}\ \emph {et~al.}(2002)\citenamefont {Harman},
  \citenamefont {Taylor}, \citenamefont {Walsh},\ and\ \citenamefont
  {LaForge}}]{ha.ta.02}%
  \BibitemOpen
  \bibfield  {author} {\bibinfo {author} {\bibfnamefont {T.~C.}\ \bibnamefont
  {Harman}}, \bibinfo {author} {\bibfnamefont {P.~J.}\ \bibnamefont {Taylor}},
  \bibinfo {author} {\bibfnamefont {M.~P.}\ \bibnamefont {Walsh}}, \ and\
  \bibinfo {author} {\bibfnamefont {B.~E.}\ \bibnamefont {LaForge}},\ }\href
  {\doibase 10.1126/science.1072886} {\bibfield  {journal} {\bibinfo  {journal}
  {Science}\ }\textbf {\bibinfo {volume} {297}},\ \bibinfo {pages} {2229}
  (\bibinfo {year} {2002})}\BibitemShut {NoStop}%
\bibitem [{\citenamefont {Chowdhury}\ \emph {et~al.}(2009)\citenamefont
  {Chowdhury}, \citenamefont {Prasher}, \citenamefont {Lofgreen}, \citenamefont
  {Chrysler}, \citenamefont {Narasimhan}, \citenamefont {Mahajan},
  \citenamefont {Koester}, \citenamefont {Alley},\ and\ \citenamefont
  {Venkatasubramanian}}]{ch.pr.09}%
  \BibitemOpen
  \bibfield  {author} {\bibinfo {author} {\bibfnamefont {I.}~\bibnamefont
  {Chowdhury}}, \bibinfo {author} {\bibfnamefont {R.}~\bibnamefont {Prasher}},
  \bibinfo {author} {\bibfnamefont {K.}~\bibnamefont {Lofgreen}}, \bibinfo
  {author} {\bibfnamefont {G.}~\bibnamefont {Chrysler}}, \bibinfo {author}
  {\bibfnamefont {S.}~\bibnamefont {Narasimhan}}, \bibinfo {author}
  {\bibfnamefont {R.}~\bibnamefont {Mahajan}}, \bibinfo {author} {\bibfnamefont
  {D.}~\bibnamefont {Koester}}, \bibinfo {author} {\bibfnamefont
  {R.}~\bibnamefont {Alley}}, \ and\ \bibinfo {author} {\bibfnamefont
  {R.}~\bibnamefont {Venkatasubramanian}},\ }\href {\doibase
  10.1038/nnano.2008.417} {\bibfield  {journal} {\bibinfo  {journal} {Nat
  Nano}\ }\textbf {\bibinfo {volume} {4}},\ \bibinfo {pages} {235} (\bibinfo
  {year} {2009})}\BibitemShut {NoStop}%
\bibitem [{\citenamefont {Zhang}\ \emph {et~al.}(2011)\citenamefont {Zhang},
  \citenamefont {Dresselhaus}, \citenamefont {Shi}, \citenamefont {Ren},\ and\
  \citenamefont {Chen}}]{zh.dr.11}%
  \BibitemOpen
  \bibfield  {author} {\bibinfo {author} {\bibfnamefont {Y.}~\bibnamefont
  {Zhang}}, \bibinfo {author} {\bibfnamefont {M.~S.}\ \bibnamefont
  {Dresselhaus}}, \bibinfo {author} {\bibfnamefont {Y.}~\bibnamefont {Shi}},
  \bibinfo {author} {\bibfnamefont {Z.}~\bibnamefont {Ren}}, \ and\ \bibinfo
  {author} {\bibfnamefont {G.}~\bibnamefont {Chen}},\ }\href {\doibase
  10.1021/nl104090j} {\bibfield  {journal} {\bibinfo  {journal} {Nano Letters}\
  }\textbf {\bibinfo {volume} {11}},\ \bibinfo {pages} {1166} (\bibinfo {year}
  {2011})}\BibitemShut {NoStop}%
\bibitem [{\citenamefont {Majumdar}(2004)}]{maju.04}%
  \BibitemOpen
  \bibfield  {author} {\bibinfo {author} {\bibfnamefont {A.}~\bibnamefont
  {Majumdar}},\ }\href {\doibase 10.1126/science.1093164} {\bibfield  {journal}
  {\bibinfo  {journal} {Science}\ }\textbf {\bibinfo {volume} {303}},\ \bibinfo
  {pages} {777} (\bibinfo {year} {2004})}\BibitemShut {NoStop}%
\bibitem [{\citenamefont {Poudel}\ \emph {et~al.}(2008)\citenamefont {Poudel},
  \citenamefont {Hao}, \citenamefont {Ma}, \citenamefont {Lan}, \citenamefont
  {Minnich}, \citenamefont {Yu}, \citenamefont {Yan}, \citenamefont {Wang},
  \citenamefont {Muto}, \citenamefont {Vashaee}, \citenamefont {Chen},
  \citenamefont {Liu}, \citenamefont {Dresselhaus}, \citenamefont {Chen},\ and\
  \citenamefont {Ren}}]{po.ha.08}%
  \BibitemOpen
  \bibfield  {author} {\bibinfo {author} {\bibfnamefont {B.}~\bibnamefont
  {Poudel}}, \bibinfo {author} {\bibfnamefont {Q.}~\bibnamefont {Hao}},
  \bibinfo {author} {\bibfnamefont {Y.}~\bibnamefont {Ma}}, \bibinfo {author}
  {\bibfnamefont {Y.}~\bibnamefont {Lan}}, \bibinfo {author} {\bibfnamefont
  {A.}~\bibnamefont {Minnich}}, \bibinfo {author} {\bibfnamefont
  {B.}~\bibnamefont {Yu}}, \bibinfo {author} {\bibfnamefont {X.}~\bibnamefont
  {Yan}}, \bibinfo {author} {\bibfnamefont {D.}~\bibnamefont {Wang}}, \bibinfo
  {author} {\bibfnamefont {A.}~\bibnamefont {Muto}}, \bibinfo {author}
  {\bibfnamefont {D.}~\bibnamefont {Vashaee}}, \bibinfo {author} {\bibfnamefont
  {X.}~\bibnamefont {Chen}}, \bibinfo {author} {\bibfnamefont {J.}~\bibnamefont
  {Liu}}, \bibinfo {author} {\bibfnamefont {M.~S.}\ \bibnamefont
  {Dresselhaus}}, \bibinfo {author} {\bibfnamefont {G.}~\bibnamefont {Chen}}, \
  and\ \bibinfo {author} {\bibfnamefont {Z.}~\bibnamefont {Ren}},\ }\href
  {\doibase 10.1126/science.1156446} {\bibfield  {journal} {\bibinfo  {journal}
  {Science}\ }\textbf {\bibinfo {volume} {320}},\ \bibinfo {pages} {634}
  (\bibinfo {year} {2008})}\BibitemShut {NoStop}%
\bibitem [{\citenamefont {Venkatasubramanian}\ \emph
  {et~al.}(2001)\citenamefont {Venkatasubramanian}, \citenamefont {Siivola},
  \citenamefont {Colpitts},\ and\ \citenamefont {O'Quinn}}]{ve.si.01}%
  \BibitemOpen
  \bibfield  {author} {\bibinfo {author} {\bibfnamefont {R.}~\bibnamefont
  {Venkatasubramanian}}, \bibinfo {author} {\bibfnamefont {E.}~\bibnamefont
  {Siivola}}, \bibinfo {author} {\bibfnamefont {T.}~\bibnamefont {Colpitts}}, \
  and\ \bibinfo {author} {\bibfnamefont {B.}~\bibnamefont {O'Quinn}},\ }\href
  {\doibase 10.1038/35098012} {\bibfield  {journal} {\bibinfo  {journal}
  {Nature}\ }\textbf {\bibinfo {volume} {413}},\ \bibinfo {pages} {597}
  (\bibinfo {year} {2001})}\BibitemShut {NoStop}%
\bibitem [{\citenamefont {Arita}\ \emph {et~al.}(2008)\citenamefont {Arita},
  \citenamefont {Kuroki}, \citenamefont {Held}, \citenamefont {Lukoyanov},
  \citenamefont {Skornyakov},\ and\ \citenamefont {Anisimov}}]{ar.ku.08}%
  \BibitemOpen
  \bibfield  {author} {\bibinfo {author} {\bibfnamefont {R.}~\bibnamefont
  {Arita}}, \bibinfo {author} {\bibfnamefont {K.}~\bibnamefont {Kuroki}},
  \bibinfo {author} {\bibfnamefont {K.}~\bibnamefont {Held}}, \bibinfo {author}
  {\bibfnamefont {A.~V.}\ \bibnamefont {Lukoyanov}}, \bibinfo {author}
  {\bibfnamefont {S.}~\bibnamefont {Skornyakov}}, \ and\ \bibinfo {author}
  {\bibfnamefont {V.~I.}\ \bibnamefont {Anisimov}},\ }\href {\doibase
  10.1103/PhysRevB.78.115121} {\bibfield  {journal} {\bibinfo  {journal} {Phys.
  Rev. B}\ }\textbf {\bibinfo {volume} {78}},\ \bibinfo {pages} {115121}
  (\bibinfo {year} {2008})}\BibitemShut {NoStop}%
\bibitem [{\citenamefont {Hsu}\ \emph {et~al.}(2004)\citenamefont {Hsu},
  \citenamefont {Loo}, \citenamefont {Guo}, \citenamefont {Chen}, \citenamefont
  {Dyck}, \citenamefont {Uher}, \citenamefont {Hogan}, \citenamefont
  {Polychroniadis},\ and\ \citenamefont {Kanatzidis}}]{hs.lo.04}%
  \BibitemOpen
  \bibfield  {author} {\bibinfo {author} {\bibfnamefont {K.~F.}\ \bibnamefont
  {Hsu}}, \bibinfo {author} {\bibfnamefont {S.}~\bibnamefont {Loo}}, \bibinfo
  {author} {\bibfnamefont {F.}~\bibnamefont {Guo}}, \bibinfo {author}
  {\bibfnamefont {W.}~\bibnamefont {Chen}}, \bibinfo {author} {\bibfnamefont
  {J.~S.}\ \bibnamefont {Dyck}}, \bibinfo {author} {\bibfnamefont
  {C.}~\bibnamefont {Uher}}, \bibinfo {author} {\bibfnamefont {T.}~\bibnamefont
  {Hogan}}, \bibinfo {author} {\bibfnamefont {E.~K.}\ \bibnamefont
  {Polychroniadis}}, \ and\ \bibinfo {author} {\bibfnamefont {M.~G.}\
  \bibnamefont {Kanatzidis}},\ }\href {\doibase 10.1126/science.1092963}
  {\bibfield  {journal} {\bibinfo  {journal} {Science}\ }\textbf {\bibinfo
  {volume} {303}},\ \bibinfo {pages} {818} (\bibinfo {year}
  {2004})}\BibitemShut {NoStop}%
\bibitem [{\citenamefont {Beyer}\ \emph {et~al.}(2002)\citenamefont {Beyer},
  \citenamefont {Nurnus}, \citenamefont {B\"ottner}, \citenamefont {Lambrecht},
  \citenamefont {Roch},\ and\ \citenamefont {Bauer}}]{be.nu.02}%
  \BibitemOpen
  \bibfield  {author} {\bibinfo {author} {\bibfnamefont {H.}~\bibnamefont
  {Beyer}}, \bibinfo {author} {\bibfnamefont {J.}~\bibnamefont {Nurnus}},
  \bibinfo {author} {\bibfnamefont {H.}~\bibnamefont {B\"ottner}}, \bibinfo
  {author} {\bibfnamefont {A.}~\bibnamefont {Lambrecht}}, \bibinfo {author}
  {\bibfnamefont {T.}~\bibnamefont {Roch}}, \ and\ \bibinfo {author}
  {\bibfnamefont {G.}~\bibnamefont {Bauer}},\ }\href@noop {} {\bibfield
  {journal} {\bibinfo  {journal} {Applied Physics Letters}\ }\textbf {\bibinfo
  {volume} {80}} (\bibinfo {year} {2002})}\BibitemShut {NoStop}%
\bibitem [{\citenamefont {Sales}, \citenamefont {Mandrus},\ and\ \citenamefont
  {Williams}(1996)}]{sa.ma.96}%
  \BibitemOpen
  \bibfield  {author} {\bibinfo {author} {\bibfnamefont {B.~C.}\ \bibnamefont
  {Sales}}, \bibinfo {author} {\bibfnamefont {D.}~\bibnamefont {Mandrus}}, \
  and\ \bibinfo {author} {\bibfnamefont {R.~K.}\ \bibnamefont {Williams}},\
  }\href {\doibase 10.1126/science.272.5266.1325} {\bibfield  {journal}
  {\bibinfo  {journal} {Science}\ }\textbf {\bibinfo {volume} {272}},\ \bibinfo
  {pages} {1325} (\bibinfo {year} {1996})},\ \Eprint
  {http://arxiv.org/abs/http://science.sciencemag.org/content/272/5266/1325.full.pdf}
  {http://science.sciencemag.org/content/272/5266/1325.full.pdf} \BibitemShut
  {NoStop}%
\bibitem [{\citenamefont {Pardo}, \citenamefont {Botana},\ and\ \citenamefont
  {Baldomir}(2013)}]{pa.bo.13}%
  \BibitemOpen
  \bibfield  {author} {\bibinfo {author} {\bibfnamefont {V.}~\bibnamefont
  {Pardo}}, \bibinfo {author} {\bibfnamefont {A.~S.}\ \bibnamefont {Botana}}, \
  and\ \bibinfo {author} {\bibfnamefont {D.}~\bibnamefont {Baldomir}},\ }\href
  {\doibase 10.1103/PhysRevB.87.125148} {\bibfield  {journal} {\bibinfo
  {journal} {Phys. Rev. B}\ }\textbf {\bibinfo {volume} {87}},\ \bibinfo
  {pages} {125148} (\bibinfo {year} {2013})}\BibitemShut {NoStop}%
\bibitem [{\citenamefont {Pardo}, \citenamefont {Botana},\ and\ \citenamefont
  {Baldomir}(2012)}]{pa.bo.13a}%
  \BibitemOpen
  \bibfield  {author} {\bibinfo {author} {\bibfnamefont {V.}~\bibnamefont
  {Pardo}}, \bibinfo {author} {\bibfnamefont {A.~S.}\ \bibnamefont {Botana}}, \
  and\ \bibinfo {author} {\bibfnamefont {D.}~\bibnamefont {Baldomir}},\ }\href
  {\doibase 10.1103/PhysRevB.86.165114} {\bibfield  {journal} {\bibinfo
  {journal} {Phys. Rev. B}\ }\textbf {\bibinfo {volume} {86}},\ \bibinfo
  {pages} {165114} (\bibinfo {year} {2012})}\BibitemShut {NoStop}%
\bibitem [{\citenamefont {Snyder}\ and\ \citenamefont
  {Toberer}(2008)}]{sn.to.08}%
  \BibitemOpen
  \bibfield  {author} {\bibinfo {author} {\bibfnamefont {G.~J.}\ \bibnamefont
  {Snyder}}\ and\ \bibinfo {author} {\bibfnamefont {E.~S.}\ \bibnamefont
  {Toberer}},\ }\href {\doibase 10.1038/nmat2090} {\bibfield  {journal}
  {\bibinfo  {journal} {Nat Mater}\ }\textbf {\bibinfo {volume} {7}},\ \bibinfo
  {pages} {105} (\bibinfo {year} {2008})}\BibitemShut {NoStop}%
\bibitem [{\citenamefont {Dresselhaus}\ \emph {et~al.}(2007)\citenamefont
  {Dresselhaus}, \citenamefont {chen}, \citenamefont {Tang}, \citenamefont
  {Yang}, \citenamefont {Lee}, \citenamefont {Wang}, \citenamefont {Ren},
  \citenamefont {Fleurial},\ and\ \citenamefont {Gogna}}]{dr.ch.07}%
  \BibitemOpen
  \bibfield  {author} {\bibinfo {author} {\bibfnamefont {M.~S.}\ \bibnamefont
  {Dresselhaus}}, \bibinfo {author} {\bibfnamefont {G.}~\bibnamefont {chen}},
  \bibinfo {author} {\bibfnamefont {M.~Y.}\ \bibnamefont {Tang}}, \bibinfo
  {author} {\bibfnamefont {R.~G.}\ \bibnamefont {Yang}}, \bibinfo {author}
  {\bibfnamefont {H.}~\bibnamefont {Lee}}, \bibinfo {author} {\bibfnamefont
  {D.~Z.}\ \bibnamefont {Wang}}, \bibinfo {author} {\bibfnamefont {Z.~F.}\
  \bibnamefont {Ren}}, \bibinfo {author} {\bibfnamefont {J.-P.}\ \bibnamefont
  {Fleurial}}, \ and\ \bibinfo {author} {\bibfnamefont {P.}~\bibnamefont
  {Gogna}},\ }\href {\doibase 10.1002/adma.200600527} {\bibfield  {journal}
  {\bibinfo  {journal} {Advanced Materials}\ }\textbf {\bibinfo {volume}
  {19}},\ \bibinfo {pages} {1043} (\bibinfo {year} {2007})}\BibitemShut
  {NoStop}%
\bibitem [{\citenamefont {Dubi}\ and\ \citenamefont
  {Di~Ventra}(2011)}]{du.ve.11}%
  \BibitemOpen
  \bibfield  {author} {\bibinfo {author} {\bibfnamefont {Y.}~\bibnamefont
  {Dubi}}\ and\ \bibinfo {author} {\bibfnamefont {M.}~\bibnamefont
  {Di~Ventra}},\ }\href {\doibase 10.1103/RevModPhys.83.131} {\bibfield
  {journal} {\bibinfo  {journal} {Rev. Mod. Phys.}\ }\textbf {\bibinfo {volume}
  {83}},\ \bibinfo {pages} {131} (\bibinfo {year} {2011})}\BibitemShut
  {NoStop}%
\bibitem [{\citenamefont {Vineis}\ \emph {et~al.}(2010)\citenamefont {Vineis},
  \citenamefont {Shakouri}, \citenamefont {Majumdar},\ and\ \citenamefont
  {Kanatzidis}}]{vi.sh.10}%
  \BibitemOpen
  \bibfield  {author} {\bibinfo {author} {\bibfnamefont {C.~J.}\ \bibnamefont
  {Vineis}}, \bibinfo {author} {\bibfnamefont {A.}~\bibnamefont {Shakouri}},
  \bibinfo {author} {\bibfnamefont {A.}~\bibnamefont {Majumdar}}, \ and\
  \bibinfo {author} {\bibfnamefont {M.~G.}\ \bibnamefont {Kanatzidis}},\ }\href
  {\doibase 10.1002/adma.201000839} {\bibfield  {journal} {\bibinfo  {journal}
  {Advanced Materials}\ }\textbf {\bibinfo {volume} {22}},\ \bibinfo {pages}
  {3970} (\bibinfo {year} {2010})}\BibitemShut {NoStop}%
\bibitem [{\citenamefont {Sootsman}, \citenamefont {Chung},\ and\ \citenamefont
  {Kanatzidis}(2009)}]{so.ch.09}%
  \BibitemOpen
  \bibfield  {author} {\bibinfo {author} {\bibfnamefont {J.~R.}\ \bibnamefont
  {Sootsman}}, \bibinfo {author} {\bibfnamefont {D.~Y.}\ \bibnamefont {Chung}},
  \ and\ \bibinfo {author} {\bibfnamefont {M.~G.}\ \bibnamefont {Kanatzidis}},\
  }\href {\doibase 10.1002/anie.200900598} {\bibfield  {journal} {\bibinfo
  {journal} {Angewandte Chemie International Edition}\ }\textbf {\bibinfo
  {volume} {48}},\ \bibinfo {pages} {8616} (\bibinfo {year}
  {2009})}\BibitemShut {NoStop}%
\bibitem [{\citenamefont {Shakouri}(2011)}]{shak.11}%
  \BibitemOpen
  \bibfield  {author} {\bibinfo {author} {\bibfnamefont {A.}~\bibnamefont
  {Shakouri}},\ }\href {\doibase 10.1146/annurev-matsci-062910-100445}
  {\bibfield  {journal} {\bibinfo  {journal} {Annual Review of Materials
  Research}\ }\textbf {\bibinfo {volume} {41}},\ \bibinfo {pages} {399}
  (\bibinfo {year} {2011})}\BibitemShut {NoStop}%
\bibitem [{\citenamefont {Dorda}\ \emph {et~al.}(2016)\citenamefont {Dorda},
  \citenamefont {Ganahl}, \citenamefont {Andergassen}, \citenamefont {von~der
  Linden},\ and\ \citenamefont {Arrigoni}}]{do.ga.16}%
  \BibitemOpen
  \bibfield  {author} {\bibinfo {author} {\bibfnamefont {A.}~\bibnamefont
  {Dorda}}, \bibinfo {author} {\bibfnamefont {M.}~\bibnamefont {Ganahl}},
  \bibinfo {author} {\bibfnamefont {S.}~\bibnamefont {Andergassen}}, \bibinfo
  {author} {\bibfnamefont {W.}~\bibnamefont {von~der Linden}}, \ and\ \bibinfo
  {author} {\bibfnamefont {E.}~\bibnamefont {Arrigoni}},\ }\href {\doibase
  10.1103/PhysRevB.94.245125} {\bibfield  {journal} {\bibinfo  {journal} {Phys.
  Rev. B}\ }\textbf {\bibinfo {volume} {94}},\ \bibinfo {pages} {245125}
  (\bibinfo {year} {2016})}\BibitemShut {NoStop}%
\bibitem [{\citenamefont {Freericks}\ and\ \citenamefont
  {Zlati\'{c}}(2007)}]{fr.zl.07}%
  \BibitemOpen
  \bibfield  {author} {\bibinfo {author} {\bibfnamefont {J.~K.}\ \bibnamefont
  {Freericks}}\ and\ \bibinfo {author} {\bibfnamefont {V.}~\bibnamefont
  {Zlati\'{c}}},\ }\href {\doibase 10.1002/pssb.200674611} {\bibfield
  {journal} {\bibinfo  {journal} {physica status solidi (b)}\ }\textbf
  {\bibinfo {volume} {244}},\ \bibinfo {pages} {2351} (\bibinfo {year}
  {2007})}\BibitemShut {NoStop}%
\bibitem [{\citenamefont {Georges}\ \emph {et~al.}(1996)\citenamefont
  {Georges}, \citenamefont {Kotliar}, \citenamefont {Krauth},\ and\
  \citenamefont {Rozenberg}}]{ge.ko.96}%
  \BibitemOpen
  \bibfield  {author} {\bibinfo {author} {\bibfnamefont {A.}~\bibnamefont
  {Georges}}, \bibinfo {author} {\bibfnamefont {G.}~\bibnamefont {Kotliar}},
  \bibinfo {author} {\bibfnamefont {W.}~\bibnamefont {Krauth}}, \ and\ \bibinfo
  {author} {\bibfnamefont {M.~J.}\ \bibnamefont {Rozenberg}},\ }\href {\doibase
  10.1103/RevModPhys.68.13} {\bibfield  {journal} {\bibinfo  {journal} {Rev.
  Mod. Phys.}\ }\textbf {\bibinfo {volume} {68}},\ \bibinfo {pages} {13}
  (\bibinfo {year} {1996})}\BibitemShut {NoStop}%
\bibitem [{\citenamefont {Vollhardt}(2010)}]{Voll.10}%
  \BibitemOpen
  \bibfield  {author} {\bibinfo {author} {\bibfnamefont {D.}~\bibnamefont
  {Vollhardt}},\ }in\ \href {\doibase 10.1063/1.3518901} {\emph {\bibinfo
  {booktitle} {Lecture Notes on the Physics of Strongly Correlated Systems}}},\
  \bibinfo {series} {AIP Conf. Proc.}, Vol.\ \bibinfo {volume} {1297},\
  \bibinfo {editor} {edited by\ \bibinfo {editor} {\bibfnamefont
  {A.}~\bibnamefont {Avella}}\ and\ \bibinfo {editor} {\bibfnamefont
  {F.}~\bibnamefont {Mancini}}}\ (\bibinfo {address} {AIP, New York},\ \bibinfo
  {year} {2010})\ pp.\ \bibinfo {pages} {339--403}\BibitemShut {NoStop}%
\bibitem [{\citenamefont {Metzner}\ and\ \citenamefont
  {Vollhardt}(1989)}]{me.vo.89}%
  \BibitemOpen
  \bibfield  {author} {\bibinfo {author} {\bibfnamefont {W.}~\bibnamefont
  {Metzner}}\ and\ \bibinfo {author} {\bibfnamefont {D.}~\bibnamefont
  {Vollhardt}},\ }\href {\doibase 10.1103/PhysRevLett.62.324} {\bibfield
  {journal} {\bibinfo  {journal} {Phys. Rev. Lett.}\ }\textbf {\bibinfo
  {volume} {62}},\ \bibinfo {pages} {324} (\bibinfo {year} {1989})}\BibitemShut
  {NoStop}%
\bibitem [{\citenamefont {Kubo}(1957)}]{kubo.57}%
  \BibitemOpen
  \bibfield  {author} {\bibinfo {author} {\bibfnamefont {R.}~\bibnamefont
  {Kubo}},\ }\href {\doibase 10.1143/JPSJ.12.570} {\bibfield  {journal}
  {\bibinfo  {journal} {Journal of the Physical Society of Japan}\ }\textbf
  {\bibinfo {volume} {12}},\ \bibinfo {pages} {570} (\bibinfo {year}
  {1957})}\BibitemShut {NoStop}%
\bibitem [{\citenamefont {Schwinger}(1961)}]{schw.61}%
  \BibitemOpen
  \bibfield  {author} {\bibinfo {author} {\bibfnamefont {J.}~\bibnamefont
  {Schwinger}},\ }\href {http://dx.doi.org/10.1063/1.1703727} {\bibfield
  {journal} {\bibinfo  {journal} {J. Math. Phys.}\ }\textbf {\bibinfo {volume}
  {2}},\ \bibinfo {pages} {407} (\bibinfo {year} {1961})}\BibitemShut {NoStop}%
\bibitem [{\citenamefont {Baym}\ and\ \citenamefont
  {Kadanoff}(1961)}]{ba.ka.61}%
  \BibitemOpen
  \bibfield  {author} {\bibinfo {author} {\bibfnamefont {G.}~\bibnamefont
  {Baym}}\ and\ \bibinfo {author} {\bibfnamefont {L.~P.}\ \bibnamefont
  {Kadanoff}},\ }\href {\doibase 10.1103/PhysRev.124.287} {\bibfield  {journal}
  {\bibinfo  {journal} {Phys. Rev.}\ }\textbf {\bibinfo {volume} {124}},\
  \bibinfo {pages} {287} (\bibinfo {year} {1961})}\BibitemShut {NoStop}%
\bibitem [{\citenamefont {Kadanoff}\ and\ \citenamefont
  {Baym}(1962)}]{kad.baym}%
  \BibitemOpen
  \bibfield  {author} {\bibinfo {author} {\bibfnamefont {L.~P.}\ \bibnamefont
  {Kadanoff}}\ and\ \bibinfo {author} {\bibfnamefont {G.}~\bibnamefont
  {Baym}},\ }\href@noop {} {\emph {\bibinfo {title} {Quantum Statistical
  Mechanics: Green's Function Methods in Equilibrium and Nonequilibrium
  Problems}}}\ (\bibinfo  {publisher} {Addison-Wesley},\ \bibinfo {address}
  {Redwood City, CA},\ \bibinfo {year} {1962})\BibitemShut {NoStop}%
\bibitem [{\citenamefont {Keldysh}(1965)}]{keld.65}%
  \BibitemOpen
  \bibfield  {author} {\bibinfo {author} {\bibfnamefont {L.~V.}\ \bibnamefont
  {Keldysh}},\ }\href
  {http://www.jetp.ac.ru/cgi-bin/e/index/e/20/4/p1018?a=list} {\bibfield
  {journal} {\bibinfo  {journal} {Sov. Phys. JETP}\ }\textbf {\bibinfo {volume}
  {20}},\ \bibinfo {pages} {1018} (\bibinfo {year} {1965})}\BibitemShut
  {NoStop}%
\bibitem [{\citenamefont {Anderson}(1961)}]{ande.61}%
  \BibitemOpen
  \bibfield  {author} {\bibinfo {author} {\bibfnamefont {P.~W.}\ \bibnamefont
  {Anderson}},\ }\href {\doibase 10.1103/PhysRev.124.41} {\bibfield  {journal}
  {\bibinfo  {journal} {Phys. Rev.}\ }\textbf {\bibinfo {volume} {124}},\
  \bibinfo {pages} {41} (\bibinfo {year} {1961})}\BibitemShut {NoStop}%
\bibitem [{\citenamefont {Arrigoni}, \citenamefont {Knap},\ and\ \citenamefont
  {von~der Linden}(2013)}]{ar.kn.13}%
  \BibitemOpen
  \bibfield  {author} {\bibinfo {author} {\bibfnamefont {E.}~\bibnamefont
  {Arrigoni}}, \bibinfo {author} {\bibfnamefont {M.}~\bibnamefont {Knap}}, \
  and\ \bibinfo {author} {\bibfnamefont {W.}~\bibnamefont {von~der Linden}},\
  }\href {\doibase 10.1103/PhysRevLett.110.086403} {\bibfield  {journal}
  {\bibinfo  {journal} {Phys. Rev. Lett.}\ }\textbf {\bibinfo {volume} {110}},\
  \bibinfo {pages} {086403} (\bibinfo {year} {2013})}\BibitemShut {NoStop}%
\bibitem [{\citenamefont {Dorda}\ \emph {et~al.}(2014)\citenamefont {Dorda},
  \citenamefont {Nuss}, \citenamefont {von~der Linden},\ and\ \citenamefont
  {Arrigoni}}]{do.nu.14}%
  \BibitemOpen
  \bibfield  {author} {\bibinfo {author} {\bibfnamefont {A.}~\bibnamefont
  {Dorda}}, \bibinfo {author} {\bibfnamefont {M.}~\bibnamefont {Nuss}},
  \bibinfo {author} {\bibfnamefont {W.}~\bibnamefont {von~der Linden}}, \ and\
  \bibinfo {author} {\bibfnamefont {E.}~\bibnamefont {Arrigoni}},\ }\href
  {\doibase 10.1103/PhysRevB.89.165105} {\bibfield  {journal} {\bibinfo
  {journal} {Phys. Rev. B}\ }\textbf {\bibinfo {volume} {89}},\ \bibinfo
  {pages} {165105} (\bibinfo {year} {2014})}\BibitemShut {NoStop}%
\bibitem [{\citenamefont {Titvinidze}\ \emph {et~al.}(2015)\citenamefont
  {Titvinidze}, \citenamefont {Dorda}, \citenamefont {von~der Linden},\ and\
  \citenamefont {Arrigoni}}]{ti.do.15}%
  \BibitemOpen
  \bibfield  {author} {\bibinfo {author} {\bibfnamefont {I.}~\bibnamefont
  {Titvinidze}}, \bibinfo {author} {\bibfnamefont {A.}~\bibnamefont {Dorda}},
  \bibinfo {author} {\bibfnamefont {W.}~\bibnamefont {von~der Linden}}, \ and\
  \bibinfo {author} {\bibfnamefont {E.}~\bibnamefont {Arrigoni}},\ }\href
  {\doibase 10.1103/PhysRevB.92.245125} {\bibfield  {journal} {\bibinfo
  {journal} {Phys. Rev. B}\ }\textbf {\bibinfo {volume} {92}},\ \bibinfo
  {pages} {245125} (\bibinfo {year} {2015})}\BibitemShut {NoStop}%
\bibitem [{\citenamefont {Dorda}\ \emph {et~al.}(2017)\citenamefont {Dorda},
  \citenamefont {Sorantin}, \citenamefont {von~der Linden},\ and\ \citenamefont
  {Arrigoni}}]{do.so.17}%
  \BibitemOpen
  \bibfield  {author} {\bibinfo {author} {\bibfnamefont {A.}~\bibnamefont
  {Dorda}}, \bibinfo {author} {\bibfnamefont {M.}~\bibnamefont {Sorantin}},
  \bibinfo {author} {\bibfnamefont {W.}~\bibnamefont {von~der Linden}}, \ and\
  \bibinfo {author} {\bibfnamefont {E.}~\bibnamefont {Arrigoni}},\ }\href
  {http://stacks.iop.org/1367-2630/19/i=6/a=063005} {\bibfield  {journal}
  {\bibinfo  {journal} {New Journal of Physics}\ }\textbf {\bibinfo {volume}
  {19}},\ \bibinfo {pages} {063005} (\bibinfo {year} {2017})}\BibitemShut
  {NoStop}%
\bibitem [{\citenamefont {Okamoto}(2007)}]{okam.07}%
  \BibitemOpen
  \bibfield  {author} {\bibinfo {author} {\bibfnamefont {S.}~\bibnamefont
  {Okamoto}},\ }\href {\doibase 10.1103/PhysRevB.76.035105} {\bibfield
  {journal} {\bibinfo  {journal} {Phys. Rev. B}\ }\textbf {\bibinfo {volume}
  {76}},\ \bibinfo {pages} {035105} (\bibinfo {year} {2007})}\BibitemShut
  {NoStop}%
\bibitem [{\citenamefont {Okamoto}(2008)}]{okam.08}%
  \BibitemOpen
  \bibfield  {author} {\bibinfo {author} {\bibfnamefont {S.}~\bibnamefont
  {Okamoto}},\ }\href {\doibase 10.1103/PhysRevLett.101.116807} {\bibfield
  {journal} {\bibinfo  {journal} {Phys. Rev. Lett.}\ }\textbf {\bibinfo
  {volume} {101}},\ \bibinfo {pages} {116807} (\bibinfo {year}
  {2008})}\BibitemShut {NoStop}%
\bibitem [{\citenamefont {Okamoto}\ and\ \citenamefont
  {Millis}(2004)}]{ok.mi.04}%
  \BibitemOpen
  \bibfield  {author} {\bibinfo {author} {\bibfnamefont {S.}~\bibnamefont
  {Okamoto}}\ and\ \bibinfo {author} {\bibfnamefont {A.~J.}\ \bibnamefont
  {Millis}},\ }\href {\doibase 10.1103/PhysRevB.70.075101} {\bibfield
  {journal} {\bibinfo  {journal} {Phys. Rev. B}\ }\textbf {\bibinfo {volume}
  {70}},\ \bibinfo {pages} {075101} (\bibinfo {year} {2004})}\BibitemShut
  {NoStop}%
\bibitem [{\citenamefont {Chen}\ and\ \citenamefont
  {Freericks}(2007)}]{ch.fr.07}%
  \BibitemOpen
  \bibfield  {author} {\bibinfo {author} {\bibfnamefont {L.}~\bibnamefont
  {Chen}}\ and\ \bibinfo {author} {\bibfnamefont {J.~K.}\ \bibnamefont
  {Freericks}},\ }\href {\doibase 10.1103/PhysRevB.75.125114} {\bibfield
  {journal} {\bibinfo  {journal} {Phys. Rev. B}\ }\textbf {\bibinfo {volume}
  {75}},\ \bibinfo {pages} {125114} (\bibinfo {year} {2007})}\BibitemShut
  {NoStop}%
\bibitem [{\citenamefont {Hale}\ and\ \citenamefont
  {Freericks}(2012)}]{ha.fr.12}%
  \BibitemOpen
  \bibfield  {author} {\bibinfo {author} {\bibfnamefont {S.~T.~F.}\
  \bibnamefont {Hale}}\ and\ \bibinfo {author} {\bibfnamefont {J.~K.}\
  \bibnamefont {Freericks}},\ }\href {\doibase 10.1103/PhysRevB.85.205444}
  {\bibfield  {journal} {\bibinfo  {journal} {Phys. Rev. B}\ }\textbf {\bibinfo
  {volume} {85}},\ \bibinfo {pages} {205444} (\bibinfo {year}
  {2012})}\BibitemShut {NoStop}%
\bibitem [{\citenamefont {Bakalov}\ \emph {et~al.}(2016)\citenamefont
  {Bakalov}, \citenamefont {Nasr~Esfahani}, \citenamefont {Covaci},
  \citenamefont {Peeters}, \citenamefont {Tempere},\ and\ \citenamefont
  {Locquet}}]{ba.es.16}%
  \BibitemOpen
  \bibfield  {author} {\bibinfo {author} {\bibfnamefont {P.}~\bibnamefont
  {Bakalov}}, \bibinfo {author} {\bibfnamefont {D.}~\bibnamefont
  {Nasr~Esfahani}}, \bibinfo {author} {\bibfnamefont {L.}~\bibnamefont
  {Covaci}}, \bibinfo {author} {\bibfnamefont {F.~M.}\ \bibnamefont {Peeters}},
  \bibinfo {author} {\bibfnamefont {J.}~\bibnamefont {Tempere}}, \ and\
  \bibinfo {author} {\bibfnamefont {J.-P.}\ \bibnamefont {Locquet}},\ }\href
  {\doibase 10.1103/PhysRevB.93.165112} {\bibfield  {journal} {\bibinfo
  {journal} {Phys. Rev. B}\ }\textbf {\bibinfo {volume} {93}},\ \bibinfo
  {pages} {165112} (\bibinfo {year} {2016})}\BibitemShut {NoStop}%
\bibitem [{\citenamefont {Charlebois}\ \emph {et~al.}(2013)\citenamefont
  {Charlebois}, \citenamefont {Hassan}, \citenamefont {Karan}, \citenamefont
  {S{\'en\'e}chal},\ and\ \citenamefont {Tremblay}}]{ch.ha.13}%
  \BibitemOpen
  \bibfield  {author} {\bibinfo {author} {\bibfnamefont {M.}~\bibnamefont
  {Charlebois}}, \bibinfo {author} {\bibfnamefont {S.~R.}\ \bibnamefont
  {Hassan}}, \bibinfo {author} {\bibfnamefont {R.}~\bibnamefont {Karan}},
  \bibinfo {author} {\bibfnamefont {D.}~\bibnamefont {S{\'en\'e}chal}}, \ and\
  \bibinfo {author} {\bibfnamefont {A.-M.~S.}\ \bibnamefont {Tremblay}},\
  }\href {\doibase 10.1103/PhysRevB.87.035137} {\bibfield  {journal} {\bibinfo
  {journal} {Phys. Rev. B}\ }\textbf {\bibinfo {volume} {87}},\ \bibinfo
  {pages} {035137} (\bibinfo {year} {2013})}\BibitemShut {NoStop}%
\bibitem [{\citenamefont {Lee}\ and\ \citenamefont
  {MacDonald}(2006)}]{le.do.06}%
  \BibitemOpen
  \bibfield  {author} {\bibinfo {author} {\bibfnamefont {W.-C.}\ \bibnamefont
  {Lee}}\ and\ \bibinfo {author} {\bibfnamefont {A.~H.}\ \bibnamefont
  {MacDonald}},\ }\href {\doibase 10.1103/PhysRevB.74.075106} {\bibfield
  {journal} {\bibinfo  {journal} {Phys. Rev. B}\ }\textbf {\bibinfo {volume}
  {74}},\ \bibinfo {pages} {075106} (\bibinfo {year} {2006})}\BibitemShut
  {NoStop}%
\bibitem [{\citenamefont {Lee}\ and\ \citenamefont
  {MacDonald}(2007)}]{le.do.07}%
  \BibitemOpen
  \bibfield  {author} {\bibinfo {author} {\bibfnamefont {W.-C.}\ \bibnamefont
  {Lee}}\ and\ \bibinfo {author} {\bibfnamefont {A.~H.}\ \bibnamefont
  {MacDonald}},\ }\href {\doibase 10.1103/PhysRevB.76.075339} {\bibfield
  {journal} {\bibinfo  {journal} {Phys. Rev. B}\ }\textbf {\bibinfo {volume}
  {76}},\ \bibinfo {pages} {075339} (\bibinfo {year} {2007})}\BibitemShut
  {NoStop}%
\bibitem [{\citenamefont {Schmidt}\ and\ \citenamefont {Monien}()}]{sc.mo.02u}%
  \BibitemOpen
  \bibfield  {author} {\bibinfo {author} {\bibfnamefont {P.}~\bibnamefont
  {Schmidt}}\ and\ \bibinfo {author} {\bibfnamefont {H.}~\bibnamefont
  {Monien}},\ }\href {http://arxiv.org/abs/cond-mat/0202046} {\enquote
  {\bibinfo {title} {Nonequilibrium dynamical mean -- field theory of a
  strongly correlated system},}\ }\bibinfo {note}
  {ArXiv:cond-mat/0202046}\BibitemShut {NoStop}%
\bibitem [{\citenamefont {Freericks}, \citenamefont {Turkowski},\ and\
  \citenamefont {Zlati{\'{c}}}(2006)}]{fr.tu.06}%
  \BibitemOpen
  \bibfield  {author} {\bibinfo {author} {\bibfnamefont {J.~K.}\ \bibnamefont
  {Freericks}}, \bibinfo {author} {\bibfnamefont {V.~M.}\ \bibnamefont
  {Turkowski}}, \ and\ \bibinfo {author} {\bibfnamefont {V.}~\bibnamefont
  {Zlati{\'{c}}}},\ }\href {\doibase 10.1103/PhysRevLett.97.266408} {\bibfield
  {journal} {\bibinfo  {journal} {Phys. Rev. Lett.}\ }\textbf {\bibinfo
  {volume} {97}},\ \bibinfo {pages} {266408} (\bibinfo {year}
  {2006})}\BibitemShut {NoStop}%
\bibitem [{\citenamefont {Freericks}(2008)}]{free.08}%
  \BibitemOpen
  \bibfield  {author} {\bibinfo {author} {\bibfnamefont {J.~K.}\ \bibnamefont
  {Freericks}},\ }\href {\doibase 10.1103/PhysRevB.77.075109} {\bibfield
  {journal} {\bibinfo  {journal} {Phys. Rev. B}\ }\textbf {\bibinfo {volume}
  {77}},\ \bibinfo {pages} {075109} (\bibinfo {year} {2008})}\BibitemShut
  {NoStop}%
\bibitem [{\citenamefont {Joura}, \citenamefont {Freericks},\ and\
  \citenamefont {Pruschke}(2008)}]{jo.fr.08}%
  \BibitemOpen
  \bibfield  {author} {\bibinfo {author} {\bibfnamefont {A.~V.}\ \bibnamefont
  {Joura}}, \bibinfo {author} {\bibfnamefont {J.~K.}\ \bibnamefont
  {Freericks}}, \ and\ \bibinfo {author} {\bibfnamefont {T.}~\bibnamefont
  {Pruschke}},\ }\href {\doibase 10.1103/PhysRevLett.101.196401} {\bibfield
  {journal} {\bibinfo  {journal} {Phys. Rev. Lett.}\ }\textbf {\bibinfo
  {volume} {101}},\ \bibinfo {pages} {196401} (\bibinfo {year}
  {2008})}\BibitemShut {NoStop}%
\bibitem [{\citenamefont {Eckstein}, \citenamefont {Kollar},\ and\
  \citenamefont {Werner}(2009)}]{ec.ko.09}%
  \BibitemOpen
  \bibfield  {author} {\bibinfo {author} {\bibfnamefont {M.}~\bibnamefont
  {Eckstein}}, \bibinfo {author} {\bibfnamefont {M.}~\bibnamefont {Kollar}}, \
  and\ \bibinfo {author} {\bibfnamefont {P.}~\bibnamefont {Werner}},\ }\href
  {\doibase 10.1103/PhysRevLett.103.056403} {\bibfield  {journal} {\bibinfo
  {journal} {Phys. Rev. Lett.}\ }\textbf {\bibinfo {volume} {103}},\ \bibinfo
  {pages} {056403} (\bibinfo {year} {2009})}\BibitemShut {NoStop}%
\bibitem [{\citenamefont {Titvinidze}\ \emph {et~al.}(2016)\citenamefont
  {Titvinidze}, \citenamefont {Dorda}, \citenamefont {von~der Linden},\ and\
  \citenamefont {Arrigoni}}]{ti.do.16}%
  \BibitemOpen
  \bibfield  {author} {\bibinfo {author} {\bibfnamefont {I.}~\bibnamefont
  {Titvinidze}}, \bibinfo {author} {\bibfnamefont {A.}~\bibnamefont {Dorda}},
  \bibinfo {author} {\bibfnamefont {W.}~\bibnamefont {von~der Linden}}, \ and\
  \bibinfo {author} {\bibfnamefont {E.}~\bibnamefont {Arrigoni}},\ }\href
  {\doibase 10.1103/PhysRevB.94.245142} {\bibfield  {journal} {\bibinfo
  {journal} {Phys. Rev. B}\ }\textbf {\bibinfo {volume} {94}},\ \bibinfo
  {pages} {245142} (\bibinfo {year} {2016})}\BibitemShut {NoStop}%
\bibitem [{\citenamefont {Dorda}, \citenamefont {Titvinidze},\ and\
  \citenamefont {Arrigoni}(2016)}]{do.ti.16}%
  \BibitemOpen
  \bibfield  {author} {\bibinfo {author} {\bibfnamefont {A.}~\bibnamefont
  {Dorda}}, \bibinfo {author} {\bibfnamefont {I.}~\bibnamefont {Titvinidze}}, \
  and\ \bibinfo {author} {\bibfnamefont {E.}~\bibnamefont {Arrigoni}},\ }\href
  {http://stacks.iop.org/1742-6596/696/i=1/a=012003} {\bibfield  {journal}
  {\bibinfo  {journal} {Journal of Physics: Conference Series}\ }\textbf
  {\bibinfo {volume} {696}},\ \bibinfo {pages} {012003} (\bibinfo {year}
  {2016})}\BibitemShut {NoStop}%
\bibitem [{\citenamefont {Haug}\ and\ \citenamefont {Jauho}(1998)}]{ha.ja}%
  \BibitemOpen
  \bibfield  {author} {\bibinfo {author} {\bibfnamefont {H.}~\bibnamefont
  {Haug}}\ and\ \bibinfo {author} {\bibfnamefont {A.-P.}\ \bibnamefont
  {Jauho}},\ }\href@noop {} {\emph {\bibinfo {title} {Quantum Kinetics in
  Transport and Optics of Semiconductors}}}\ (\bibinfo  {publisher}
  {Springer},\ \bibinfo {address} {Heidelberg},\ \bibinfo {year}
  {1998})\BibitemShut {NoStop}%
\bibitem [{\citenamefont {Rammer}\ and\ \citenamefont
  {Smith}(1986)}]{ra.sm.86}%
  \BibitemOpen
  \bibfield  {author} {\bibinfo {author} {\bibfnamefont {J.}~\bibnamefont
  {Rammer}}\ and\ \bibinfo {author} {\bibfnamefont {H.}~\bibnamefont {Smith}},\
  }\href {\doibase 10.1103/RevModPhys.58.323} {\bibfield  {journal} {\bibinfo
  {journal} {Rev. Mod. Phys.}\ }\textbf {\bibinfo {volume} {58}},\ \bibinfo
  {pages} {323} (\bibinfo {year} {1986})}\BibitemShut {NoStop}%
\bibitem [{Gre()}]{Green_g}%
  \BibitemOpen
  \href@noop {} {}\bibinfo {note} {By $\underline{g}_{r}(\omega,k_\perp)$ we
  denote the decoupled ($v_r=0$) surface Green's function for the right lead.
  In general, the decoupled Green's function of the decoupled right lead
  depends on $k_\perp$ and two $z$ coordinates, e.g.
  $\underline{g}_{r}(\omega,k_\perp,z,z')$. The surface Green's function at the
  edge layer is then defined as ${\underline g}_{r}(\omega, {\vv
  k})=\underline{g}_{r}(\omega,k_\perp,z=1,z'=1)$, where $z=1$ denotes the
  layer adjacent to the correlated one. The knowledge of
  $\underline{g}_{r/l}(\omega,k_\perp)$ is sufficient to calculate the
  properties of the correlated layer, we don't need this Green's function for
  the other values of $z,z'$. Notice that $\underline{g}_{\alpha}$ is different
  from $\underline{G}_{\alpha}$, which describes the corresponding Green's
  function when the leads and the correlated layer are coupled to each other
  (i.e. $v_l=v_r \neq 0$). $\underline{G}_{\alpha}$ are non-equilibrium Green's
  function, while $g_{\alpha}$ are in equilibrium. Therefore, one can use the
  fluctuation dissipation theorem to calculate the Keldish Green's functions
  $g_{\alpha}^K$ from the retarded Green's function $g_{\alpha}^R$, but of
  course not for the $\underline{G}$.}\BibitemShut {Stop}%
\bibitem [{\citenamefont {Breuer}\ and\ \citenamefont
  {Petruccione}(2009)}]{br.pe}%
  \BibitemOpen
  \bibfield  {author} {\bibinfo {author} {\bibfnamefont {H.-P.}\ \bibnamefont
  {Breuer}}\ and\ \bibinfo {author} {\bibfnamefont {F.}~\bibnamefont
  {Petruccione}},\ }\href@noop {} {\emph {\bibinfo {title} {The Theory of Open
  Quantum Systems}}}\ (\bibinfo  {publisher} {Oxford University Press},\
  \bibinfo {address} {Oxford, England},\ \bibinfo {year} {2009})\BibitemShut
  {NoStop}%
\bibitem [{\citenamefont {Schaller}(2014)}]{scha}%
  \BibitemOpen
  \bibfield  {author} {\bibinfo {author} {\bibfnamefont {G.}~\bibnamefont
  {Schaller}},\ }\href@noop {} {\emph {\bibinfo {title} {Open Quantum Systems
  Far from Equilibrium}}},\ Lecture notes in physics\ (\bibinfo  {publisher}
  {Springer},\ \bibinfo {address} {Heidelberg},\ \bibinfo {year}
  {2014})\BibitemShut {NoStop}%
\bibitem [{\citenamefont {Nuss}\ \emph {et~al.}(2015)\citenamefont {Nuss},
  \citenamefont {Dorn}, \citenamefont {Dorda}, \citenamefont {von~der Linden},\
  and\ \citenamefont {Arrigoni}}]{nu.do.15}%
  \BibitemOpen
  \bibfield  {author} {\bibinfo {author} {\bibfnamefont {M.}~\bibnamefont
  {Nuss}}, \bibinfo {author} {\bibfnamefont {G.}~\bibnamefont {Dorn}}, \bibinfo
  {author} {\bibfnamefont {A.}~\bibnamefont {Dorda}}, \bibinfo {author}
  {\bibfnamefont {W.}~\bibnamefont {von~der Linden}}, \ and\ \bibinfo {author}
  {\bibfnamefont {E.}~\bibnamefont {Arrigoni}},\ }\href {\doibase
  10.1103/PhysRevB.92.125128} {\bibfield  {journal} {\bibinfo  {journal} {Phys.
  Rev. B}\ }\textbf {\bibinfo {volume} {92}},\ \bibinfo {pages} {125128}
  (\bibinfo {year} {2015})}\BibitemShut {NoStop}%
\bibitem [{\citenamefont {Schwarz}\ \emph {et~al.}(2016)\citenamefont
  {Schwarz}, \citenamefont {Goldstein}, \citenamefont {Dorda}, \citenamefont
  {Arrigoni}, \citenamefont {Weichselbaum},\ and\ \citenamefont {von
  Delft}}]{sc.go.16}%
  \BibitemOpen
  \bibfield  {author} {\bibinfo {author} {\bibfnamefont {F.}~\bibnamefont
  {Schwarz}}, \bibinfo {author} {\bibfnamefont {M.}~\bibnamefont {Goldstein}},
  \bibinfo {author} {\bibfnamefont {A.}~\bibnamefont {Dorda}}, \bibinfo
  {author} {\bibfnamefont {E.}~\bibnamefont {Arrigoni}}, \bibinfo {author}
  {\bibfnamefont {A.}~\bibnamefont {Weichselbaum}}, \ and\ \bibinfo {author}
  {\bibfnamefont {J.}~\bibnamefont {von Delft}},\ }\href {\doibase
  10.1103/PhysRevB.94.155142} {\bibfield  {journal} {\bibinfo  {journal} {Phys.
  Rev. B}\ }\textbf {\bibinfo {volume} {94}},\ \bibinfo {pages} {155142}
  (\bibinfo {year} {2016})}\BibitemShut {NoStop}%
\bibitem [{\citenamefont {Jauho}(2006)}]{jauh}%
  \BibitemOpen
  \bibfield  {author} {\bibinfo {author} {\bibfnamefont {A.-P.}\ \bibnamefont
  {Jauho}},\ }\href {https://nanohub.org/resources/1877} {\  (\bibinfo {year}
  {2006})},\ \bibinfo {note} {https://nanohub.org/resources/1877}\BibitemShut
  {NoStop}%
\bibitem [{\citenamefont {Meir}\ and\ \citenamefont
  {Wingreen}(1992)}]{me.wi.92}%
  \BibitemOpen
  \bibfield  {author} {\bibinfo {author} {\bibfnamefont {Y.}~\bibnamefont
  {Meir}}\ and\ \bibinfo {author} {\bibfnamefont {N.~S.}\ \bibnamefont
  {Wingreen}},\ }\href {http://link.aps.org/abstract/PRL/v68/p2512} {\bibfield
  {journal} {\bibinfo  {journal} {Phys. Rev. Lett.}\ }\textbf {\bibinfo
  {volume} {68}},\ \bibinfo {pages} {2512} (\bibinfo {year}
  {1992})}\BibitemShut {NoStop}%
\bibitem [{\citenamefont {Ren}\ \emph {et~al.}(2012)\citenamefont {Ren},
  \citenamefont {Zhu}, \citenamefont {Gubernatis}, \citenamefont {Wang},\ and\
  \citenamefont {Li}}]{re.zh.12}%
  \BibitemOpen
  \bibfield  {author} {\bibinfo {author} {\bibfnamefont {J.}~\bibnamefont
  {Ren}}, \bibinfo {author} {\bibfnamefont {J.-X.}\ \bibnamefont {Zhu}},
  \bibinfo {author} {\bibfnamefont {J.~E.}\ \bibnamefont {Gubernatis}},
  \bibinfo {author} {\bibfnamefont {C.}~\bibnamefont {Wang}}, \ and\ \bibinfo
  {author} {\bibfnamefont {B.}~\bibnamefont {Li}},\ }\href {\doibase
  10.1103/PhysRevB.85.155443} {\bibfield  {journal} {\bibinfo  {journal} {Phys.
  Rev. B}\ }\textbf {\bibinfo {volume} {85}},\ \bibinfo {pages} {155443}
  (\bibinfo {year} {2012})}\BibitemShut {NoStop}%
\bibitem [{Ell()}]{Elliptic_K}%
  \BibitemOpen
  \href@noop {} {}\bibinfo {note} {We use the definition of ${\cal K}$,
  implemented in Mathematica and Matlab ${\cal K}(x) = \int_{0}^{1}
  [(1-t^{2})(1- x t^{2})] ^{-1/2}\;dt$.}\BibitemShut {Stop}%
\bibitem [{\citenamefont {Ouyang}\ \emph {et~al.}(2016)\citenamefont {Ouyang},
  \citenamefont {Xie}, \citenamefont {Zhang}, \citenamefont {Peng},\ and\
  \citenamefont {Chen}}]{ou.yu.16}%
  \BibitemOpen
  \bibfield  {author} {\bibinfo {author} {\bibfnamefont {Y.}~\bibnamefont
  {Ouyang}}, \bibinfo {author} {\bibfnamefont {Y.}~\bibnamefont {Xie}},
  \bibinfo {author} {\bibfnamefont {Z.}~\bibnamefont {Zhang}}, \bibinfo
  {author} {\bibfnamefont {Q.}~\bibnamefont {Peng}}, \ and\ \bibinfo {author}
  {\bibfnamefont {Y.}~\bibnamefont {Chen}},\ }\href {\doibase
  10.1063/1.4972831} {\bibfield  {journal} {\bibinfo  {journal} {Journal of
  Applied Physics}\ }\textbf {\bibinfo {volume} {120}},\ \bibinfo {pages}
  {235109} (\bibinfo {year} {2016})},\ \Eprint
  {http://arxiv.org/abs/http://dx.doi.org/10.1063/1.4972831}
  {http://dx.doi.org/10.1063/1.4972831} \BibitemShut {NoStop}%
\bibitem [{Gap()}]{Gap}%
  \BibitemOpen
  \href@noop {} {}\bibinfo {note} {The isolated correlated layer for $U=12$ is
  in the Mott insulator phase and the spectral function has a pronounced gap.
  Due to the coupling to the leads the system always remains metallic. The
  spectral function at $\omega=0$ should have a very narrow Kondo peak.
  However, to resolve this peak in AMEA one needs $N_B \simeq 15$, which at the
  moment is too time consuming in a self-consistent DMFT calculation.On the
  other hand, for $T>T_{K}$ our results are still reliable.}\BibitemShut
  {Stop}%
\bibitem [{Not()}]{Note_appendix}%
  \BibitemOpen
  \href@noop {} {}\bibinfo {note} {Note that in Ref. \cite{okam.07} the current
  is calculated per spin and per unit area, while in our case summation over
  spin is included.}\BibitemShut {Stop}%
\bibitem [{\citenamefont {Niu}, \citenamefont {Lin},\ and\ \citenamefont
  {Lin}(1999)}]{niu_equation_1999}%
  \BibitemOpen
  \bibfield  {author} {\bibinfo {author} {\bibfnamefont {C.}~\bibnamefont
  {Niu}}, \bibinfo {author} {\bibfnamefont {D.~L.}\ \bibnamefont {Lin}}, \ and\
  \bibinfo {author} {\bibfnamefont {T.~H.}\ \bibnamefont {Lin}},\ }\href
  {http://iopscience.iop.org/0953-8984/11/6/015} {\bibfield  {journal}
  {\bibinfo  {journal} {Journal of Physics: Condensed Matter}\ }\textbf
  {\bibinfo {volume} {11}},\ \bibinfo {pages} {1511} (\bibinfo {year}
  {1999})}\BibitemShut {NoStop}%
\end{thebibliography}
 
%

\end{document}